# Inter-Subject Analysis: Inferring Sparse Interactions with Dense Intra-Graphs

Cong Ma[*]    Junwei Lu[*]    Han Liu[*]

**Abstract**

We develop a new modeling framework for Inter-Subject Analysis (ISA). The goal of ISA is to explore the dependency structure between different subjects with the intra-subject dependency as nuisance. It has important applications in neuroscience to explore the functional connectivity between brain regions under natural stimuli. Our framework is based on the Gaussian graphical models, under which ISA can be converted to the problem of estimation and inference of the inter-subject precision matrix. The main statistical challenge is that we do not impose sparsity constraint on the whole precision matrix and we only assume the inter-subject part is sparse. For estimation, we propose to estimate an alternative parameter to get around the non-sparse issue and it can achieve asymptotic consistency even if the intra-subject dependency is dense. For inference, we propose an "untangle and chord" procedure to de-bias our estimator. It is valid without the sparsity assumption on the inverse Hessian of the log-likelihood function. This inferential method is general and can be applied to many other statistical problems, thus it is of independent theoretical interest. Numerical experiments on both simulated and brain imaging data validate our methods and theory.

# 1   Introduction

Inter-Subject Analysis (ISA) refers to the inference of the dependency structures between different subjects with intra-subject dependencies as nuisance. The subject may be a pathway consisting of an assembly of genes or a group of stocks from the same sector in financial markets. Often the dependency structure between different subjects is of scientific interest while the dependencies within each subject are complicated and hard to infer. The goal of ISA is to explore the inter-subject dependencies with intra-subject dependencies as nuisance.

[*]Department of Operations Research and Financial Engineering, Princeton University, Princeton, NJ 08544, USA; Email: {congm, junweil, hanliu}@princeton.edu

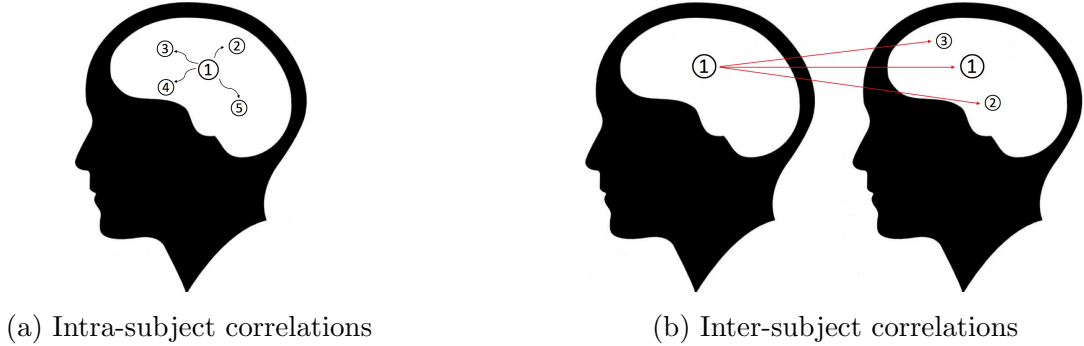

(a) Intra-subject correlations  (b) Inter-subject correlations

Figure 1: (a) illustrates the intra-subject correlations where the correlations among voxels in the same subject are calculated. (b) illustrates the inter-subject correlations where the correlations of voxels are computed across subjects.

## 1.1 Motivating Example

To motivate the use of Inter-Subject Analysis (ISA), we consider the functional magnetic resonance imaging (fMRI) data analysis. fMRI provides scientists a noninvasive way to observe the neural activity in the human brain (Ashby, 2011). Traditionally, fMRI measurements are obtained under highly controlled experimental settings where subjects are asked to perform identical and demanding attention tasks. Recent studies show that neuronal responses and brain activities are more reliable under naturalistic stimuli, for instance, watching a movie episode or listening to an audiobook (Mechler et al., 1998; Yao et al., 2007; Belitski et al., 2008). This motivates the use of fMRI under more naturalistic settings (Zacks et al., 2001; Hartley et al., 2003; Bartels and Zeki, 2004). However, this brings substantial noise to the fMRI measurements since intrinsic cognitive processes that are not related to the ongoing stimuli cannot be constrained or removed as in controlled research settings (Hasson et al., 2004; Simony et al., 2016). Conventional intra-subject analysis which computes voxel-by-voxel correlations in the same subject can be influenced by such noise and fail to detect the stimulus-induced correlations.

Hasson et al. (2004) introduced Inter-subject correlations (ISC) to partially resolve this problem. Instead of computing the intra-subject correlations, ISC calculates the correlation coefficients of corresponding voxels across different experimental subjects. It is based on the assumption that individual variations are uncorrelated across subjects and high inter-subject correlations indicate stimulus related activations. Although ISC can isolate the individual noise, as a measure of marginal dependence, it fails to eliminate the confounding effects of other factors (Horwitz and Rapoport, 1988; Lee et al., 2011). Conditional dependence has long been studied to remedy this problem in both statistics and biology community (Marrelec et al., 2006; Huang et al., 2009; Varoquaux et al., 2012).

## 1.2 Modeling Framework

In this paper, we propose a new modeling framework named Inter-Subject Analysis (ISA) to infer the conditional dependency between different subjects. Formally, let $X = (X_1, \ldots, X_d)^\top \sim N(0, \Sigma^*)$ be a $d$-dimensional Gaussian random vector with precision matrix $\Omega^* = (\omega^*_{jk})$. Let $\mathcal{G}_1$ and $\mathcal{G}_2$ be two disjoint subsets of $\{1, 2, \ldots, d\}$ with cardinality $|\mathcal{G}_1| = d_1$ and $|\mathcal{G}_2| = d_2 := d - d_1$. We use $X_{\mathcal{G}_1}$ and $X_{\mathcal{G}_2}$ to denote the corresponding subvectors of $X$ and they represent features of two different subjects. We use $\Sigma_1^*$, $\Sigma_2^*$ and $\Sigma_{12}^*$ to denote the covariance within $X_{\mathcal{G}_1}$, within $X_{\mathcal{G}_2}$ and between $X_{\mathcal{G}_1}$ and $X_{\mathcal{G}_2}$ respectively. The submatrices $\Omega_1^*$, $\Omega_2^*$ and $\Omega_{12}^*$ are defined similarly. It is well known that $X_j$ and $X_k$ are conditionally independent given the remaining variables if and only if $\omega^*_{jk} = 0$. Our modeling assumption is that $\Omega_{12}^*$ purely represents the dependency driven by the common stimuli while $\Omega_1^*$ and $\Omega_2^*$ can be influenced by the individual cognitive process. Hence we are willing to assume that $\Omega_{12}^*$ is sparse while reluctant to put any sparsity constraint on $\Omega_1^*$ or $\Omega_2^*$. The main statistical challenge we address in this paper is to obtain estimation consistency and valid inference of $\Omega_{12}^*$ under non-sparse nuisance parameters in the high dimensional setting.

## 1.3 Contributions

There are two major contributions of this paper.

Our first contribution is a new estimator for $\Omega_{12}^*$ when the precision matrix $\Omega^*$ is not sparse. The idea is to find an alternative sparse matrix to estimate. The candidate we propose is

$$\Theta^* := \Omega^* - (\Sigma_{\mathcal{G}}^*)^{-1}, \tag{1.1}$$

where $\Sigma_{\mathcal{G}}^* = \text{diag}(\Sigma_1^*, \Sigma_2^*)$ and $\text{diag}(A, B)$ denotes the diagonal block matrix whose diagonals are $A$ and $B$. The key observation is that if $\Omega_{12}^*$ is sparse, then $\Theta^*$ is also sparse. More precisely, we have $\|\Theta^*\|_0 \leq 2s^2 + 2s$ whenever $\|\Omega_{12}^*\|_0 \leq s$, where $\|A\|_0$ counts the number of non-zero entries in $A$. This observation holds true even if both $\Omega_1^*$ and $\Omega_2^*$ are dense. We illustrate this phenomenon using a numerical example in Figure 2. Following this observation, we can reparametrize the precision matrix using $\Omega^* = \Theta^* + (\Sigma_{\mathcal{G}}^*)^{-1}$, in which $\Theta^*$ contains the parameter of interest (as $\Theta^*$'s off-diagonal block $\Theta_{12}^* = \Omega_{12}^*$) and $(\Sigma_{\mathcal{G}}^*)^{-1}$ is a nuisance parameter. The estimator we propose named **S**parse edge es**T**imato**R** for **I**ntense **N**uisance **G**raph**S** (STRINGS) has the following form:

$$\widehat{\Theta} = \arg\min \text{Tr}(\Theta \widehat{\Sigma}) - \log |\widehat{\Sigma}_{\mathcal{G}} \Theta \widehat{\Sigma}_{\mathcal{G}} + \widehat{\Sigma}_{\mathcal{G}}| + \lambda \|\Theta\|_{1,1},$$

where $\widehat{\Sigma}$ and $\widehat{\Sigma}_{\mathcal{G}}$ are plug-in estimators for $\Sigma^*$ and $\Sigma_{\mathcal{G}}^*$ and $\|A\|_{1,1}$ is the element-wise $\ell_1$ norm of $A$. The consistency for the STRINGS estimator can be achieved under mild conditions.

Our second contribution is to propose a general "untangle and chord" procedure to de-bias high dimensional estimators when the inverse Hessian of the log-likelihood function is not sparse. In general, a de-biased estimator $\widehat{\beta}^u$ takes the following form:

$$\widehat{\beta}^u = \widehat{\beta} - M \nabla \mathcal{L}_n(\widehat{\beta}),$$

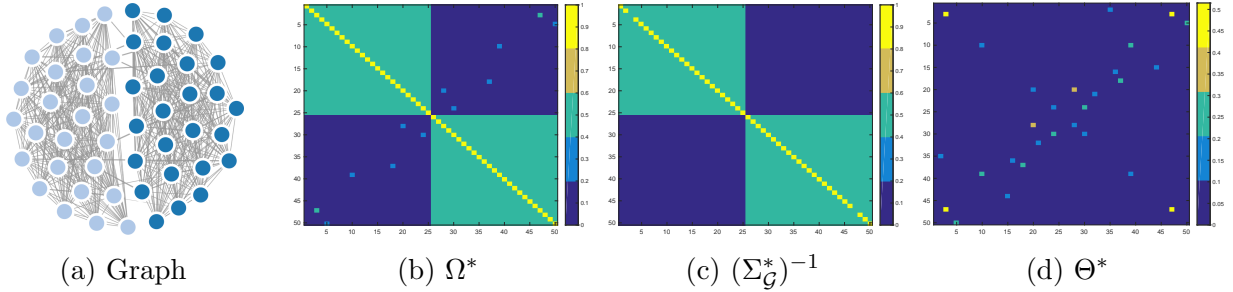

(a) Graph  (b) $\Omega^*$  (c) $(\Sigma_{\mathcal{G}}^*)^{-1}$  (d) $\Theta^*$

Figure 2: (a) shows a Gaussian graphical model with two subjects, where each color represents a subject. It can be seen that the inter-subject connections are sparse while the connections within each subject are dense. (b) shows the heatmap of the precision matrix of the Gaussian graphical model. (c) show the heatmap of the corresponding $(\Sigma_{\mathcal{G}}^*)^{-1}$. (d) shows the heatmap of $\Theta^* = \Omega^* - (\Sigma_{\mathcal{G}}^*)^{-1}$. As can be seen, $\Theta^*$ is sparse even if both $\Omega_1^*$ and $\Omega_2^*$ are dense.

where $\widehat{\beta}$ is the regularized estimator for the parameter of interest $\beta^*$, $M$ is a bias correction matrix and $\mathcal{L}_n$ denotes the negative log-likelihood function. By the Taylor expansion of $\mathcal{L}_n(\widehat{\beta})$ at the true parameter $\beta^*$, we have

$$\sqrt{n} \cdot (\widehat{\beta}^u - \beta^*) \approx -\sqrt{n} M \nabla \mathcal{L}_n(\beta^*) - \sqrt{n}[M \nabla^2 \mathcal{L}_n(\beta^*) - I](\widehat{\beta} - \beta^*).$$

Clearly, the leading term $\sqrt{n} M \nabla \mathcal{L}_n(\beta^*)$ is asymptotically normal under mild conditions. And if the remainder term $\sqrt{n}[M \nabla^2 \mathcal{L}_n(\beta^*) - I](\widehat{\beta} - \beta^*)$ converges to 0 in probability, we can conclude that $\widehat{\beta}^u$ is asymptotically normal. One way to achieve this goal is to let $M$ be a consistent estimator of the inverse of $\nabla^2 \mathcal{L}_n(\beta^*)$. This is why previous inferential methods require the sparsity assumption on the inverse Hessian.

It will be shown that the Hessian in ISA is not necessarily sparse. To get around this issue, two crucial observations are needed. First, it suffices to use constrained optimization to control the statistical error of $M \nabla^2 \mathcal{L}_n(\beta^*) - I$. Second, to prevent $M$ from tangling with $\nabla \mathcal{L}_n(\beta^*)$ and sabotaging the asymptotic normality of the leading term, we can split the data into two parts: the untangle step estimates the parameter on the first split and the chord step constructs $M$ using the second split. We show that the "untangle and chord" procedure de-biases the STRINGS estimator and we can construct valid tests and confidence intervals based on the de-biased estimator. The "untangle and chord" strategy is general and can be applied to many other high dimensional problems without the sparsity assumption on the inverse Hessian.

### 1.4 Related Work

Estimators for the precision matrix including the maximum likelihood estimator and the column-wise estimator have been considered in (Yuan and Lin, 2007; Banerjee et al., 2008; Rothman et al., 2008; Friedman et al., 2008; Ravikumar et al., 2011; Meinshausen and Bühlmann, 2006; Yuan, 2010; Cai et al., 2011). All of them require the sparsity of the whole precision matrix. Hence they are not applicable to ISA.

Inferential methods based on inverting KKT conditions (Zhang and Zhang, 2014; Van de Geer et al., 2014) and decorrelated score functions (Ning and Liu, 2014) have been extended to Gaussian graphical models (Jankova et al., 2015; Gu et al., 2015). They all require the inverse Hessian of the log-likelihood function to be sparse and hence cannot be applied in our setting. One exception is the inference for the high dimensional linear model proposed by Javanmard and Montanari (2014). Their result heavily depends on the special structure of the linear model. First, the design matrix is independent of the noise and second, the inverse Hessian matrix is simply the inverse covariance of the design, which is irrelevant to the regression parameters. Their method is difficult to extend to general estimators.

Efforts have also been made to relax the sparsity assumption on the precision matrix in Gaussian graphical estimation. Yuan and Zhang (2014) propose a decoupling approach to estimate $\Omega_{12}^*$. However, their method requires at least one of the diagonal blocks $\Omega_1^*$ or $\Omega_2^*$ to be sparse. And it is no longer valid if both of them are dense. Liu et al. (2015) shares the similar goal with ours. They proposed a density ratio framework to estimate the dependency between two groups of variables. First, their estimation theory does not apply to Gaussian distributions due to the unboundedness of the density ratio. Second, their procedure relies on approximating the normalization function using two sample U-statistics which are complicated and computationally expensive. Compared with the above works, our work not only considers the estimation consistency but also propose valid procedures to assess uncertainty in the high dimensional setting.

Besides the work mentioned above, Kolar et al. (2014) and Xia et al. (2016) focus on testing whether the inter-subject dependency matrix is identically zero. However, our procedure can give element-wise estimation and inference.

## 1.5 Notation

The following notations are used throughout the paper. For any $n \in \mathbb{N}$ we use the shorthand notation $[n] = \{1, \ldots, n\}$. For a vector $v = (v_1, \ldots, v_d)^\top \in \mathbb{R}^d$, let $\|v\|_q = (\sum_{i=1}^d v_i^q)^{1/q}, 1 \leq q < \infty$. Furthermore, let $\|v\|_\infty = \max_j |v_j|$. For a matrix $A = (A_{jk}) \in \mathbb{R}^{m \times n}$, we define $\text{supp}(A) = \{(j,k) | A_{jk} \neq 0\}$. We use $A_{j*}$ and $A_{*k}$ to denote the $j$-th row and $k$-th column of $A$ respectively. We use $\|A\|_q = \sup_{\|x\|_q=1} \|Ax\|_q$ to denote the induced $\ell_q$-norm of a matrix. In particular, $\|A\|_1 = \max_{1 \leq k \leq n} \sum_{j=1}^m |A_{jk}|$, which is the maximum absolute column sum of the matrix $A$. $\|A\|_2$ is the largest singular value of $A$. $\|A\|_\infty = \max_{1 \leq j \leq m} \sum_{k=1}^n |A_{jk}|$, which is the maximum absolute row sum of the matrix $A$. We also use $\|A\|_{\max} = \max_{jk} |A_{jk}|$, $\|A\|_{1,1} = \sum_{jk} |A_{jk}|$ and $\|A\|_F = (\sum_{jk} A_{jk}^2)^{1/2}$ to denote the $\ell_{\max}$-, $\ell_{1,1}$- and $\ell_F$-norms of the matrix $A$. $\lambda_{\min}(A)$ is used to denote the minimum eigenvalue of the matrix $A$ and $|A|$ is used to denote the determinant of $A$. We use $\Phi(x)$ to denote the cumulative distribution function of a standard normal random variable. For a sequence of random variables $X_n$, we write $X_n \rightsquigarrow X$, for some random variable $X$, if $X_n$ converges in distribution to $X$.

## 1.6 Organization of the Paper

The rest of the paper is organized as follows. In Section 2, we introduce the STRINGS estimator for the inter-subject precision matrix. In Section 3, we provide the "untangle and chord" procedure to de-bias the STRINGS estimator. In Section 4, we show theoretical results on the statistical rate of convergence for the STRINGS estimator and the asymptotic normality of the de-biased estimator. In Section 5, we demonstrate the validity of our estimation and inferential methods through numerical experiments on both simulated and brain imaging data. In Section 6, we discuss extensions of our methods and theories to non-Gaussian data and multi-subjects settings.

## 2 The STRINGS Estimator

In this section, we present the STRINGS estimator for the inter-subject precision matrix $\Omega_{12}^*$. The basic idea is to use the maximum likelihood principle. Given a data matrix $\mathbb{X} \in \mathbb{R}^{n \times d}$, where rows of $\mathbb{X}$ represent i.i.d. samples from a Gaussian distribution with mean 0 and covariance $\Sigma^*$, the negative log-likelihood for the precision matrix is given by

$$\mathcal{L}(\Omega) = \mathrm{Tr}(\Omega \widehat{\Sigma}) - \log |\Omega|, \tag{2.1}$$

where $\widehat{\Sigma} = (1/n) \cdot \mathbb{X}^\top \mathbb{X}$ is the sample covariance matrix.

Since our focus is to estimate the inter-subject dependency $\Omega_{12}^*$, a naive reparametrization of the precision matrix is $\Omega^* = \Omega_D^* + \Omega_O^*$, where $\Omega_D^*$ is the block diagonal matrix corresponding to $X_{\mathcal{G}_1}$ and $X_{\mathcal{G}_2}$, i.e., $\Omega_D^* = \mathrm{diag}(\Omega_1^*, \Omega_2^*)$. And $\Omega_O^*$ is the off-diagonal parts involving $\Omega_{12}^*$ and $\Omega_{12}^{*\top}$. Under such a reparametrization, we can reformulate (2.1) as

$$\mathcal{L}(\Omega_O, \Omega_D) = \mathrm{Tr}[(\Omega_O + \Omega_D)\widehat{\Sigma}] - \log |\Omega_O + \Omega_D|. \tag{2.2}$$

Adopting the maximum likelihood principle, we want to minimize the negative log-likelihood (2.2) with respect to the parameter $\Omega_O$. Hence we can ignore the terms independent of $\Omega_O$ in (2.2). This gives us an equivalent minimization of $\mathrm{Tr}(\Omega_O \widehat{\Sigma}) - \log |\Omega_O + \Omega_D|$ with respect to $\Omega_O$. However, the objective function still depends on the nuisance parameter $\Omega_D$ and it is difficult to obtain an estimator for $\Omega_D$. Thus this naive reparametrization will not work.

Recall that if $\Omega_{12}^*$ is $s$-sparse, then $\Theta^* = \Omega^* - (\Sigma_{\mathcal{G}}^*)^{-1}$ is $(2s^2 + 2s)$-sparse. Based on this key observation, we reparametrize the precision matrix using $\Theta^*$ and $(\Sigma_{\mathcal{G}}^*)^{-1}$, in which $\Theta^*$ contains the parameter of interest (as $\Theta_{12}^* = \Omega_{12}^*$) and $(\Sigma_{\mathcal{G}}^*)^{-1}$ is the nuisance parameter. Under the new reparametrization $\Omega^* = \Theta^* + (\Sigma_{\mathcal{G}}^*)^{-1}$, we can rewrite (2.1) as

$$\mathcal{L}(\Theta, \Sigma_{\mathcal{G}}^{-1}) = \mathrm{Tr}[(\Theta + \Sigma_{\mathcal{G}}^{-1})\widehat{\Sigma}] - \log |\Theta + \Sigma_{\mathcal{G}}^{-1}|. \tag{2.3}$$

To separate $\Theta$ from $\Sigma_{\mathcal{G}}^{-1}$, we further decompose (2.3) into the following form:

$$\mathcal{L}(\Theta, \Sigma_{\mathcal{G}}^{-1}) = \mathrm{Tr}(\Theta \widehat{\Sigma}) + \mathrm{Tr}(\Sigma_{\mathcal{G}}^{-1}\widehat{\Sigma}) - \log|\Sigma_{\mathcal{G}}^{-1}| - \log|\Sigma_{\mathcal{G}} \Theta \Sigma_{\mathcal{G}} + \Sigma_{\mathcal{G}}| - \log|\Sigma_{\mathcal{G}}^{-1}|. \tag{2.4}$$

Ignoring the terms independent of $\Theta$ in (2.4), we have that minimizing (2.4) with respect to $\Theta$ is equivalent to minimizing $\text{Tr}(\Theta\widehat{\Sigma}) - \log|\Sigma_{\mathcal{G}}\Theta\Sigma_{\mathcal{G}} + \Sigma_{\mathcal{G}}|$ with respect to $\Theta$. Now we still have the nuisance parameter $\Sigma_{\mathcal{G}}$. However in this case we can use the naive plug-in estimator for $\Sigma_{\mathcal{G}}$, i.e., $\widehat{\Sigma}_{\mathcal{G}}$ is the block diagonal matrix of $\widehat{\Sigma}$ corresponding to $X_{\mathcal{G}_1}$ and $X_{\mathcal{G}_2}$. This gives us the following empirical loss function for $\Theta$:

$$\mathcal{L}_n(\Theta) = \text{Tr}(\Theta\widehat{\Sigma}) - \log|\widehat{\Sigma}_{\mathcal{G}}\Theta\widehat{\Sigma}_{\mathcal{G}} + \widehat{\Sigma}_{\mathcal{G}}|. \tag{2.5}$$

Correspondingly, we will use $\mathcal{L}(\Theta) = \text{Tr}(\Theta\Sigma^*) - \log|\Sigma^*_{\mathcal{G}}\Theta\Sigma^*_{\mathcal{G}} + \Sigma^*_{\mathcal{G}}|$ to denote the population loss function. Since $\Theta^*$ is sparse, we further impose a sparsity penalty on the objective function. Here we choose the $\ell_{1,1}$ penalty, and the STRINGS estimator has the following form

$$\widehat{\Theta} = \arg\min \mathcal{L}_n(\Theta) + \lambda\|\Theta\|_{1,1}. \tag{2.6}$$

Note that for the empirical loss function $\mathcal{L}_n(\Theta)$ in (2.5) to be convex, $\widehat{\Sigma}_{\mathcal{G}}$ needs to be positive definite. Otherwise, the log-determinant term will always be $-\infty$. However, in the high dimensional regime, the naive plug-in estimator $\widehat{\Sigma}_{\mathcal{G}}$ will be rank deficient. To resolve this issue, we can perturb our plug-in estimator with $\sqrt{\log d/n} \cdot I$. This perturbation trick has also been used to solve the initialization problem in Cai et al. (2011). We choose the size of the perturbation to be $\sqrt{\log d/n}$ so that it will not affect the concentration property of the estimator $\widehat{\Sigma}_{\mathcal{G}}$. Although (2.6) is a convex program, solving it in high dimensions using semidefinite programming is both time-consuming and memory-consuming. We propose a computationally efficient algorithm based on alternating direction method of multipliers (ADMM) to solve (2.6). The details are deferred to Appendix Section B.

In the end of this section, we want to comment on why the block-wise regression cannot be applied to estimate $\Omega^*_{12}$. Since neighborhood selection (Meinshausen and Bühlmann, 2006; Yuan, 2010) can be used to estimate the sparse precision matrix, a natural generalization for the inter-subject precision matrix is to consider a block-wise regression method. Namely, we regress one group of variables $X_{\mathcal{G}_1}$ on the other group of variables $X_{\mathcal{G}_2}$. However, this would fail since the sparsity pattern of $\Omega^*_{21}$ cannot be translated into the sparsity pattern of the regression coefficient matrix. More precisely, based on the formula for conditional Gaussian distribution, we have

$$X_{\mathcal{G}_1}|X_{\mathcal{G}_2} = x_{\mathcal{G}_2} \sim N(\Sigma^*_{12}\Sigma^{*-1}_2 x_{\mathcal{G}_2}, \Sigma^*_1 - \Sigma^*_{12}\Sigma^{*-1}_2\Sigma^{*\top}_{12}).$$

Thus the regression function between $X_{\mathcal{G}_1}$ and $X_{\mathcal{G}_2}$ is given by $X_{\mathcal{G}_1} = \beta^\top X_{\mathcal{G}_2} + \epsilon$, where $\beta = \Sigma^{*-1}_2 \Sigma^{*\top}_{12}$ and $\epsilon \sim N(0, \Sigma^*_1 - \Sigma^*_{12}\Sigma^{*-1}_2\Sigma^{*\top}_{12})$. Then by the formula for matrix inversion in block form, we have $\Omega^*_{12} = -(\Sigma^*_1 - \Sigma^*_{12}\Sigma^{*-1}_2\Sigma^{*\top}_{12})^{-1}\Sigma^*_{12}\Sigma^{*-1}_2$. Hence, we have the following relationship between the precision matrix $\Omega^*_{12}$ and the regression coefficients $\beta$:

$$\Omega^{*\top}_{12}(\Sigma^*_1 - \Sigma^*_{12}\Sigma^{*-1}_2\Sigma^{*\top}_{12}) = -\beta \in \mathbb{R}^{d_2 \times d_1}.$$

The reason why we can use neighborhood selection to recover the precision matrix in traditional Gaussian graphical model lies crucially in the fact that $|\mathcal{G}_1| = 1$. In this situation, we have $\Sigma_1^* - \Sigma_{12}^*\Sigma_2^{*-1}\Sigma_{12}^{*\top}$ is a (positive) real number. Hence the sparsity pattern of $\beta$ is exactly the same as the sparsity pattern in $\Omega_{12}^*$. However, this does not hold when $|\mathcal{G}_1| > 1$. In such case, $\Sigma_1^* - \Sigma_{12}^*\Sigma_2^{*-1}\Sigma_{12}^{*\top}$ is a $d_1$-by-$d_1$ matrix. When $\Omega_{12}^*$ is $s$-sparse, $\beta$ is $s$-row sparse, i.e., $\beta$ has $s$ non-zero rows. However, the rate for estimating such a row sparse matrix is $\|\widehat{\beta} - \beta\|_F^2 = \mathcal{O}_{\mathbb{P}}(sd/n)$ (Negahban and Wainwright, 2011; Rohde et al., 2011). Thus no consistent estimator exists in the high dimensional setting when $d > n$.

## 3 "Untangle and Chord" the STRINGS

In this section, we introduce our proposed method to test the existence of certain inter-subject interaction and construct a confidence interval for entries in $\Omega_{12}^*$. Formally, for $1 \leq j \leq d_1$ and $d_1 + 1 \leq k \leq d$, we are interested in the following two types of inferential problems:

- Confidence Interval: For a particular parameter $\theta_{jk}^*$, where $\theta_{jk}^*$ is the $(j,k)$-th entry of $\Theta^*$, how to construct a confidence interval for it?

- Hypothesis testing: Consider the null hypothesis $H_0 : \theta_{jk}^* = 0$, how to construct a valid test for $H_0$?

To address these two types of questions, we rely on obtaining an asymptotically normal estimator of $\theta_{jk}^*$. After this, constructing confidence intervals and testing hypotheses follow naturally. Hence in the following we introduce our way to de-bias the STRINGS estimator.

As we mentioned in the introduction, KKT-inversion type of methods (Zhang and Zhang, 2014; Van de Geer et al., 2014) cannot be applied here since they require the inverse Hessian to be sparse. Recall that the population loss function is given by $\mathcal{L}(\Theta) = \text{Tr}(\Theta\Sigma^*) - \log|\Sigma_\mathcal{G}^*\Theta\Sigma_\mathcal{G}^* + \Sigma_\mathcal{G}^*|$. Its Hessian can be calculated as following:

$$\nabla^2 \mathcal{L}(\Theta) = [(\Sigma_\mathcal{G}^*\Theta + I)^{-1}\Sigma_\mathcal{G}^*] \otimes [(\Sigma_\mathcal{G}^*\Theta + I)^{-1}\Sigma_\mathcal{G}^*].$$

Thus we have $[\nabla^2 \mathcal{L}(\Theta^*)]^{-1} = \Omega^* \otimes \Omega^*$. We can see that the inverse Hessian can be dense since we do not impose any assumption on $\Omega^*$. Getting around with this difficulty requires new sets of inferential tools. Rather than inverting the KKT conditions, we propose to de-bias the STRINGS estimator in (2.6) utilizing the estimating equation for $\Theta^*$. Moreover, sample splitting is adopted to achieve the desired asymptotic normality. To see this, recall the definition of $\Theta^*$ that $\Theta^* = \Omega^* - (\Sigma_\mathcal{G}^*)^{-1}$, we can derive the following estimating equation:

$$\Sigma^*\Theta^*\Sigma_\mathcal{G}^* + \Sigma^* - \Sigma_\mathcal{G}^* = 0. \tag{3.1}$$

We first present a heuristic explanation on our de-baising procedure. Based on the sample version of (3.1), we construct a de-biased estimator as following

$$\widehat{\Theta}^u = \widehat{\Theta} - M(\widehat{\Sigma}\widehat{\Theta}\widehat{\Sigma}_\mathcal{G} + \widehat{\Sigma} - \widehat{\Sigma}_\mathcal{G})P^\top, \tag{3.2}$$

where $M$ and $P$ are two bias correction matrices to be specified later. To gain the intuition why $\widehat{\Theta}^u$ defined in (3.2) is an asymptotically normal estimator, we calculate the difference between $\widehat{\Theta}^u$ and $\Theta^*$ as follows.

$$\widehat{\Theta}^u - \Theta^* = \widehat{\Theta} - \Theta^* - M[\widehat{\Sigma}(\widehat{\Theta} - \Theta^* + \Theta^*)\widehat{\Sigma}_{\mathcal{G}} + \widehat{\Sigma} - \widehat{\Sigma}_{\mathcal{G}}]P^\top$$
$$= -M(\widehat{\Sigma}\Theta^*\widehat{\Sigma}_{\mathcal{G}} + \widehat{\Sigma} - \widehat{\Sigma}_{\mathcal{G}})P^\top + \widehat{\Theta} - \Theta^* - M\widehat{\Sigma}(\widehat{\Theta} - \Theta^*)\widehat{\Sigma}_{\mathcal{G}}P^\top.$$

Through some algebra, we have $\widehat{\Theta}^u - \Theta^* = \text{Leading} + \text{Remainder}$, where

$$\text{Leading} = -M[(\widehat{\Sigma} - \Sigma^*)(I + \Theta^*\Sigma_{\mathcal{G}}^*) - (I - \Sigma^*\Theta^*)(\widehat{\Sigma}_{\mathcal{G}} - \Sigma_{\mathcal{G}}^*)]P^\top, \tag{3.3}$$

$$\text{Remainder} = -M(\widehat{\Sigma} - \Sigma^*)\Theta^*(\widehat{\Sigma}_{\mathcal{G}} - \Sigma_{\mathcal{G}}^*)P^\top + \widehat{\Theta} - \Theta^* - M\widehat{\Sigma}(\widehat{\Theta} - \Theta^*)\widehat{\Sigma}_{\mathcal{G}}P^\top. \tag{3.4}$$

First, in order to make the Remainder term in (3.4) small, it requires $M\widehat{\Sigma} \approx I$ and $P\widehat{\Sigma}_{\mathcal{G}} \approx I$. In other words, $M$ and $P$ should function as the inverse of $\Sigma^*$ and $\Sigma_{\mathcal{G}}^*$ respectively. Second, regarding the Leading term, we can see that it is a empirical process type quantity. It is asymptotically normal provided that $M$ and $P$ are independent of the remaining random quantities in (3.3). This motivates us to utilize sample splitting to obtain the two bias correction matrices $M$ and $P$. In all, we have an "untangle and chord" procedure to de-bias the STRINGS estimator. Concretely, we split the data $\mathbb{X} \in \mathbb{R}^{2n \times d}$ into two parts $\mathcal{D}_1$ and $\mathcal{D}_2$ with equal number of samples. Note that here we inflate the sample size to $2n$. This is purely for simplifying the notations. The untangle step uses the first data $\mathcal{D}_1$ to get an initial STRINGS estimator $\widehat{\Theta}$. The chord step utilizes the second data $\mathcal{D}_2$ to obtain the bias correction matrices $M$ and $P$ with desired properties, i.e., $M\widehat{\Sigma} \approx I$ and $P\widehat{\Sigma}_{\mathcal{G}} \approx I$. Precisely we use a CLIME-type procedure to get $M$ and $P$. For $M \in \mathbb{R}^{d \times d}$, we solve the following convex program:

$$\min \quad \|M\|_\infty \tag{3.5}$$
$$\text{subject to} \quad \|M\widehat{\Sigma}' - I\|_{\max} \le \lambda',$$

where $\widehat{\Sigma}'$ is the sample covariance matrix of the second sample $\mathcal{D}_2$ and $\lambda'$ is the approximation error we want to achieve. $\widehat{\Sigma}'$ and $\lambda'$ can be viewed as two inputs to this CLIME-type procedure. We solve a similar convex problem to obtain $P \in \mathbb{R}^{d \times d}$ with different inputs and an additional block constraint:

$$\min \quad \|P\|_\infty \tag{3.6}$$
$$\text{subject to} \quad \|P\widehat{\Sigma}_{\mathcal{G}}' - I\|_{\max} \le \lambda',$$
$$P = \text{diag}(P_1, P_2), P_1 \in \mathbb{R}^{d_1 \times d_1}, P_2 \in \mathbb{R}^{d_2 \times d_2},$$

where $\widehat{\Sigma}_{\mathcal{G}}'$ is the block diagonal sample covariance matrix corresponding to $X_{\mathcal{G}_1}$ and $X_{\mathcal{G}_2}$ on the second sample $\mathcal{D}_2$. Notice here we add another constraint that $P$ needs to be a block diagonal matrix. This is expectable since $\widehat{\Sigma}_{\mathcal{G}}'$ is a block diagonal matrix. The overall de-biasing procedure is presented in Algorithm 1.

Given the de-biased estimator $\widehat{\Theta}^u = (\widehat{\theta}_{jk}^u)$ in (3.2), we can obtain a confidence interval for $\theta_{jk}^*$ and conduct hypothesis testing on $H_0 : \theta_{jk}^* = 0$ under a valid estimation of the asymptotic variance of $\widehat{\theta}_{jk}^u$. The asymptotic variance is involved, and we defer the details to Section 4.2.

**Algorithm 1** "Untangle and Chord" the STRINGS

    **Input**: $\mathbb{X} \in \mathbb{R}^{2n \times d}$, where rows of $\mathbb{X}$ represent i.i.d samples from $N(0, \Sigma^*)$.
    **Output**: $\widehat{\Theta}^u$, the de-biased estimator for $\Theta^*$.
    **Data Splitting**: Split the sample into two parts $\mathcal{D}_1$ and $\mathcal{D}_2$.
    **Covariance Estimation**: Let $\widehat{\Sigma}$ and $\widehat{\Sigma}'$ be the sample covariance matrices on $\mathcal{D}_1$ and $\mathcal{D}_2$ respectively. And $\widehat{\Sigma}_{\mathcal{G}}$ and $\widehat{\Sigma}'_{\mathcal{G}}$ are block diagonal matrices of $\widehat{\Sigma}$ and $\widehat{\Sigma}'$.
    **Untangle Step**: Get the STRINGS estimator $\widehat{\Theta}$ by Algorithm 2 using $\mathcal{D}_1$.
    **Chord Step**: Given the sample covariance matrices on $\mathcal{D}_2$, i.e., $\widehat{\Sigma}'$ and $\widehat{\Sigma}'_{\mathcal{G}}$, choose $M$ to be the minimizer of (3.5) and choose $P$ to be the minimizer of (3.6).
    **Debias**: Define the debiased estimator as

$$\widehat{\Theta}^u = \widehat{\Theta} - M(\widehat{\Sigma}\widehat{\Theta}\widehat{\Sigma}_{\mathcal{G}} + \widehat{\Sigma} - \widehat{\Sigma}_{\mathcal{G}})P^\top.$$

# 4 Theoretical Results

In this section, we provide theoretical properties of the proposed estimators. In Section 4.1, we establish the rate of convergence of the STRINGS estimator in (2.6). In Section 4.2, we prove the asymptotic normality of the de-biased estimator in (3.2). We also construct asymptotically valid tests and confidence intervals for low dimensional parameters in $\Theta^*_{12}$.

## 4.1 Estimation Consistency

In the next theorem, we give the convergence rate for the STRINGS estimator in (2.6).

**Theorem 4.1.** Suppose the inter-subject dependencies $\|\Omega^*_{12}\|_0 \leq s$ and hence $\|\Theta^*\|_0 \leq s^* := 2s^2 + 2s$. Further we assume $\|\Sigma^*\|_{\max} \leq K$ and $\|\Theta^*\|_{1,1} \leq R$ for some absolute constants $K$ and $R$. Under the sample size condition that $s^{*2}\sqrt{\log d/n} = o(1)$, there exists a constant $C > 0$ such that for sufficiently large $n$, if $\lambda = 2C\sqrt{\log d/n}$, with probability at least $1 - 4d^{-1}$, we have

$$\|\widehat{\Theta} - \Theta^*\|_F \leq \frac{14C}{\rho^2_{\min}}\sqrt{\frac{s^* \log d}{n}} \quad \text{and} \quad \|\widehat{\Theta} - \Theta^*\|_{1,1} \leq \frac{56C}{\rho^2_{\min}} s^* \sqrt{\frac{\log d}{n}},$$

provided that $\lambda_{\min}(\Sigma^*) \geq \rho_{\min} > 0$.

    A few remarks on the assumptions are in order. First, $\|\Omega^*_{12}\|_0 \leq s$ is the main assumption for our theoretical results. It imposes sparsity on $\Omega^*_{12}$, i.e., the dependency structure between $X_{\mathcal{G}_1}$ and $X_{\mathcal{G}_2}$ is sparse. Note that we do not make any assumption about the sparsity of the overall precision matrix $\Omega^*$. It can be rather dense. Second, $\|\Sigma^*\|_{\max} \leq K$ and $\|\Theta^*\|_{1,1} \leq R$ are two regularity conditions on the Gaussian graphical model. The first specifies that the covariance between any two variables cannot be too large. It is weaker than $\|\Sigma^*\|_2 \leq K$ since $\|\Sigma^*\|_{\max} \leq \|\Sigma^*\|_2$. This assumption can be commonly found in literatures on covariance

and precision matrix estimation (Bickel and Levina, 2008; Rothman et al., 2008). An easy consequence is that $\|\Sigma^*_{\mathcal{G}}\|_{\max} \leq K$ since $\Sigma^*_{\mathcal{G}}$ is the block diagonal of $\Sigma^*$. $\|\Theta^*\|_{1,1} \leq R$ requires the inter-subject dependency has constant sparsity. Similar conditions can be found in literatures on differential networks (Zhao et al., 2014).

Theorem 4.1 shares the same spirit with the convergence results for $\ell_1$ regularized maximum likelihood estimator (Rothman et al., 2008), that is the rate for estimating an $s^*$-sparse parameter $\Theta^*$ is $\sqrt{s^* \log d/n}$ in Frobenius norm. However, there are two things worth to be noted here. The first is that in Theorem 4.1, $s^*$ can be replaced with any upper bound of $\|\Theta^*\|_0$ and the result is still valid. We know $s^*$ is an upper bound of $\|\Theta^*\|_0$. In the worst case, $\|\Theta^*\|_0$ can be as large as $s^*$ and when $\|\Theta^*\|_0$ is smaller, the rate in Theorem 4.1 can be improved. Second, recall that $s^* \asymp s^2$, where $s$ is the sparsity of $\Omega^*_{12}$. Considering our goal is to estimate $\Omega^*_{12}$, the rate seems to be sub-optimal. Especially in the case $d_1 = 1$, neighborhood selection (Meinshausen and Bühlmann, 2006; Yuan, 2010) and CLIME (Cai et al., 2011) can obtain the optimal rate $\sqrt{s \log d/n}$ for the Frobenius norm. However, as we pointed out in Section 1, these methods cannot be applied when $d_1 \asymp d_2$ due to the violation of the sparsity assumption on $\Omega^*$.

## 4.2 Asymptotic Inference

In this section, we give the limiting distribution of the de-biased estimator in (3.2). The asymptotic normality result is presented in Theorem 4.5. Based on this, we propose valid asymptotic confidence intervals and test statistics for parameters in $\Theta^*_{12}$.

We first state a version of asymptotic normality result which involves population quantities.

**Theorem 4.2.** (Asymptotic Normality) Suppose the conditions in Theorem 4.1 hold. Further assume $\|\Omega^*\|_1 \leq L$ for some absolute constant $L > 0$. Let $\widehat{\Theta}^u$ be the de-biased estimator with $\lambda' = C'\sqrt{\log d/n}$, where $C'$ is a sufficiently large constant. For any $1 \leq j \leq d_1$ and $d_1 + 1 \leq k \leq d$, define the asymptotic variance as

$$\xi_{jk}^2 = (M_{j*}\Sigma^* M_{j*}^\top)[P_{k*}(I + \Sigma^*_{\mathcal{G}}\Theta^*)\Sigma^*_{\mathcal{G}}P_{k*}^\top] + (M_{j*}\Sigma^*_{\mathcal{G}}P_{k*}^\top)^2 - (M_{j*}\Sigma^* P_{k*}^\top)^2 \\ - [M_{j*}(I - \Sigma^*\Theta^*)\Sigma^*_{\mathcal{G}_2}(I - \Theta^*\Sigma^*)M_{j*}^\top](P_{k*}\Sigma^*_{\mathcal{G}}P_{k*}^\top), \quad (4.1)$$

where $\Sigma^*_{\mathcal{G}_2} = \text{diag}(0, \Sigma^*_2)$. Under the scaling condition $s^* \log d/\sqrt{n} = o(1)$, we have

$$\sqrt{n} \cdot (\widehat{\theta}^u_{jk} - \theta^*_{jk})/\xi_{jk} \rightsquigarrow N(0, 1).$$

**Remark 4.3.** $\|\Omega^*\|_1 \leq L$ is a milder condition than the sparsity constraints on $\Omega^*$ in the sense that $\Omega^*$ can be rather dense. And this is the case for ISA. To further understand the essence of this assumption, we discuss connections between $\lambda_{\min}(\Sigma^*) \geq \rho_{\min}$ and $\|\Omega^*\|_1 \leq L$. Since $\Omega^* = (\Sigma^*)^{-1}$, it is not hard to see that $\lambda_{\max}(\Omega^*) \leq 1/\rho_{\min}$. Hence we have $\max_{j \in [d]} \|\Omega^*_{*j}\|_2 \leq 1/\rho_{\min}$. Here, instead of the column-wise $\ell_2$-norm boundedness, we assume that $\max_{j \in [d]} \|\Omega^*_{*j}\|_1 \leq L$. It is indeed stronger than the $\ell_2$ one, but it is weaker than the sparsity assumption on $\Omega^*$. Moreover, as shown by the lower bound in Cai et al. (2016), imposing this assumption does not make it possible to consistently estimate the parameter. Based on Theorem 1.1 in Cai et al. (2016), we have that the optimal rate for the matrix $\ell_1$-norm is $\mathbb{E}\|\widehat{\Omega} - \Omega^*\|_1^2 \asymp d^2 \log d/n$, which means no consistent estimator for the whole precision matrix $\Omega^*$ exists when $d > n$.

To obtain the formula for the asymptotic variance $\xi_{jk}^2$ in (4.1), we use the Isserlis' theorem (Isserlis, 1916) to calculate the fourth order moment of the Gaussian distribution. We can see that $\xi_{jk}^2$ still depends on population quantities $\Sigma^*$ and $\Theta^*$. Thus $\xi_{jk}^2$ is unknown in practice, and we need to get a consistent estimator $\widehat{\xi}_{jk}^2$ to construct confidence intervals for $\theta_{jk}^*$.

**Lemma 4.4.** (Variance Estimation) Define $\widehat{\Sigma}_{\mathcal{G}_2} = \mathrm{diag}(0, \widehat{\Sigma}_2)$. For any $1 \leq j \leq d_1$ and $d_1 + 1 \leq k \leq d$, let $\widehat{\xi}_{jk}^2$ be the empirical version of (4.1), i.e.,

$$\widehat{\xi}_{jk}^2 = (M_{j*}\widehat{\Sigma}M_{j*}^\top)[P_{k*}(I + \widehat{\Sigma}_{\mathcal{G}}\widehat{\Theta})\widehat{\Sigma}_{\mathcal{G}}P_{k*}^\top] + (M_{j*}\widehat{\Sigma}_{\mathcal{G}}P_{k*}^\top)^2 - (M_{j*}\widehat{\Sigma}P_{k*}^\top)^2$$
$$- [M_{j*}(I - \widehat{\Sigma}\widehat{\Theta})\widehat{\Sigma}_{\mathcal{G}_2}(I - \widehat{\Theta}\widehat{\Sigma})M_{j*}^\top](P_{k*}\widehat{\Sigma}_{\mathcal{G}}P_{k*}^\top). \tag{4.2}$$

Then under the conditions in Theorem 4.2, $\widehat{\xi}_{jk}/\xi_{jk}$ converges in probability to 1.

Combining Theorem 4.2 and Lemma 4.4 with Slutsky's theorem, we can obtain the final version of the asymptotic normality result which does not involve any population quantity.

**Theorem 4.5.** Suppose the conditions in Theorem 4.2 hold. Let $\widehat{\Theta}^u$ be the de-biased estimator with $\lambda' = C'\sqrt{\log d/n}$, where $C'$ is a sufficiently large constant. For any $1 \leq j \leq d_1$ and $d_1 + 1 \leq k \leq d$, under the scaling condition $s^* \log d/\sqrt{n} = o(1)$, we have

$$\sqrt{n} \cdot (\widehat{\theta}_{jk}^u - \theta_{jk}^*)/\widehat{\xi}_{jk} \rightsquigarrow N(0, 1).$$

Applying Theorem 4.5, it is easy to construct asymptotically valid confidence intervals and test functions. For any $1 \leq j \leq d_1$ and $d_1 + 1 \leq k \leq d$ and the significance level $\alpha \in (0, 1)$, let

$$I_{jk}(\alpha) = \left[\widehat{\theta}_{jk}^u - \delta(\alpha, n), \widehat{\theta}_{jk}^u + \delta(\alpha, n)\right], \quad \text{where} \quad \delta(\alpha, n) = \frac{\widehat{\xi}_{jk}}{\sqrt{n}}\Phi^{-1}\left(1 - \frac{\alpha}{2}\right). \tag{4.3}$$

Also for the null hypothesis $H_0 : \theta_{jk}^* = 0$, we can construct the following test function

$$T_{jk}(\alpha) = \begin{cases} 1 & \text{if } |\sqrt{n} \cdot \widehat{\theta}_{jk}^u/\widehat{\xi}_{jk}| > \Phi^{-1}(1 - \alpha/2), \\ 0 & \text{if } |\sqrt{n} \cdot \widehat{\theta}_{jk}^u/\widehat{\xi}_{jk}| \leq \Phi^{-1}(1 - \alpha/2), \end{cases} \tag{4.4}$$

where $\alpha \in (0, 1)$ is the significance level of the test. The following Corollary proves the validity of the confidence interval and the test function.

**Corollary 4.6.** Suppose the conditions in Theorem 4.5 hold. The confidence interval in (4.3) is asymptotically valid and the type I error of (4.4) is asymptotically $\alpha$, i.e.,

$$\lim_{n \to \infty} \mathbb{P}(\theta_{jk}^* \in I_{jk}(\alpha)) = 1 - \alpha \quad \text{and} \quad \lim_{n \to \infty} \mathbb{P}_{\theta_{jk}^*=0}(T_{jk}(\alpha) = 1) = \alpha.$$

Table 1: Mean (Standard Error) of different metrics for STRINGS and GLASSO when $s = 10$

| | Precision | | Recall | | F-score | |
|---|---|---|---|---|---|---|
| $d$ | STRINGS | GLASSO | STRINGS | GLASSO | STRINGS | GLASSO |
| 30 | **0.37**(0.11) | 0.22(0.06) | **1.00**(0.00) | 1.00(0.00) | **0.53**(0.11) | 0.36(0.08) |
| 60 | **0.31**(0.09) | 0.22(0.06) | **1.00**(0.00) | 1.00(0.00) | **0.47**(0.11) | 0.36(0.07) |
| 100 | **0.31**(0.10) | 0.22(0.06) | **1.00**(0.02) | 1.00(0.02) | **0.47**(0.11) | 0.36(0.08) |
| 250 | **0.21**(0.07) | 0.18(0.06) | **0.99**(0.02) | 0.99(0.03) | **0.34**(0.10) | 0.30(0.08) |

## 5 Numerical Experiments

In this section, we conduct numerical experiments on both simulated and real data to validate our STRINGS estimator and the "untangle and chord" procedure. We also compare our method with $\ell_1$ regularized maximum likelihood estimator (GLASSO).

### 5.1 Simulated Data

For each dimension $d$ and sparsity $s$, we generate the precision matrix $\Omega^*$ as following. First, we split the variables $[d]$ into two groups $\mathcal{G}_1 = \{1, \ldots, d/2\}$ and $\mathcal{G}_2 = \{d/2+1, \ldots, d\}$ with equal cardinality. For the intra-subject precision matrices, we let $\Omega_1^* = \Omega_2^*$ to be the all 1 matrix. For the inter-subject precision matrix $\Omega_{12}^*$, we uniformly sample $s$ indices from $\mathcal{G}_1 \times \mathcal{G}_2$ to be the support of $\Omega_{12}^*$. And the corresponding values are set to be 0.5. Further $\delta I$ is added to $\Omega^*$ to make its condition number equal to $d$. Finally the precision matrix is standardized so that the diagonal entries of $\Omega^*$ are all 1's.

Under this model, we generate $n = 100$ training samples from the multivariate normal distribution with mean 0 and covariance $\Sigma^* = (\Omega^*)^{-1}$, and an independent sample of size 100 with the same distribution to choose the tuning parameter $\lambda$ in (2.6). The tuning parameter has the form $\lambda = C\sqrt{\log d/n}$ according to Theorem 4.1, where $C$ is set to be 50 uniform values across $[0, 5]$. We evaluate the estimators on the validation sample using the estimation equation in (3.1), i.e., we define the validation loss incurred by the estimator $\widehat{\Theta}$ to be

$$\mathcal{L}_{val}(\widehat{\Theta}) = \|\widehat{\Sigma}^{val}\widehat{\Theta}\widehat{\Sigma}_\mathcal{G}^{val} + \widehat{\Sigma}^{val} - \widehat{\Sigma}_\mathcal{G}^{val}\|_F, \quad (5.1)$$

where $\widehat{\Sigma}^{val}$ and $\widehat{\Sigma}_\mathcal{G}^{val}$ are the empirical covariance matrices on the validation sample. The final tuning parameter $\lambda$ is chosen to be the one with the smallest validation loss in (5.1).

#### 5.1.1 Estimation Quality

Since we are estimating the support of $\Omega_{12}^*$, we adopt standard metrics including Precision, Recall and F-score to measure its performance. They are defined as follows

$$\text{Precision} = \frac{\text{TP}}{\text{TP} + \text{FP}}, \quad \text{Recall} = \frac{\text{TP}}{\text{TP} + \text{FN}}, \quad \text{F-score} = \frac{2 \cdot \text{Precision} \cdot \text{Recall}}{\text{Precision} + \text{Recall}},$$

Table 2: Mean (Standard Error) of different metrics for STRINGS and GLASSO when $s = 30$

|  | Precision | | Recall | | F-score | |
| --- | --- | --- | --- | --- | --- | --- |
| $d$ | STRINGS | GLASSO | STRINGS | GLASSO | STRINGS | GLASSO |
| 30  | **0.43**(0.07) | 0.34(0.05) | 0.89(0.06) | **0.94**(0.05) | **0.58**(0.06) | 0.49(0.06) |
| 60  | **0.51**(0.10) | 0.25(0.04) | **1.00**(0.01) | 1.00(0.01) | **0.67**(0.08) | 0.40(0.04) |
| 100 | **0.47**(0.09) | 0.23(0.03) | **1.00**(0.01) | 1.00(0.01) | **0.63**(0.08) | 0.37(0.05) |
| 250 | **0.38**(0.08) | 0.19(0.03) | 0.99(0.02) | **1.00**(0.01) | **0.54**(0.08) | 0.32(0.04) |

Table 3: Mean (Standard Error) of different metrics for STRINGS and GLASSO when $s = 50$

|  | Precision | | Recall | | F-score | |
| --- | --- | --- | --- | --- | --- | --- |
| $d$ | STRINGS | GLASSO | STRINGS | GLASSO | STRINGS | GLASSO |
| 30  | **0.48**(0.05) | 0.40(0.04) | 0.78(0.07) | **0.86**(0.05) | **0.60**(0.04) | 0.54(0.04) |
| 60  | **0.38**(0.06) | 0.33(0.04) | 0.81(0.06) | **0.86**(0.06) | **0.51**(0.05) | 0.47(0.04) |
| 100 | **0.61**(0.09) | 0.23(0.02) | **1.00**(0.01) | 1.00(0.00) | **0.75**(0.06) | 0.38(0.03) |
| 250 | **0.50**(0.06) | 0.20(0.03) | **1.00**(0.00) | 1.00(0.00) | **0.67**(0.05) | 0.33(0.04) |

where TP is the number of true positives (nonzero entries in $\Theta_{12}^*$ are considered to be positives), FP is the false positives and FN stands for the false negatives. By definition, high Precision means the algorithm recovers substantially more truth than negatives, and high Recall means the algorithm returns most of the truth. Hence, Precision measures the exactness or quality of support recovery, and Recall is a measure of the algorithm's completeness or quantity. F-score, as the harmonic mean of Precision and Recall, combines these two into a single metric. The higher the F-score, the better the algorithm is in support recovery. For $\widehat{\Theta}_{12}$, absolute values above $1 \times 10^{-4}$ are considered to be non-zeros since we set the optimization accuracy to be $1 \times 10^{-4}$. We consider different values of $d \in \{30, 60, 100, 250\}$ and $s \in \{10, 30, 50\}$. For each configuration, we replicate for 100 times and report the mean and standard error of Precision, Recall and F-score in Tables 1 - 3. We can see that the STRINGS estimator outperforms GLASSO uniformly over all configurations of $(d, s)$. The improvement is more significant when the dimension $d$ and the sparsity $s$ are larger. This is expectable since GLASSO is not tailored for estimation under dense precision matrices. Due to the dense intra-subject precision matrices, GLASSO tends to under-regularize, which results in too many false positives. And this leads to poor Precision and F-score.

### 5.1.2 Inference Quality

For inference, we only report the results for STRINGS since GLASSO performs poorly in terms of estimation. We generate another sample of size 100 to de-bias the initial STRINGS estimator. Guided by Theorem 4.2, the tuning parameter $\lambda'$ in CLIME-type procedure is chosen to be $0.5\sqrt{\log d/n}$. By Corollary 4.6, the $(1 - \alpha) \times 100\%$ asymptotic confidence

Table 4: Average coverage probabilities and average lengths over $S$ and $S^c$ for 100 replications

| $d$ | $s$ | Avgcov$_S$ | Avgcov$_{S^c}$ | Avglen$_S$ | Avglen$_{S^c}$ |
|---|---|---|---|---|---|
| 30 | 10 | 0.9430 | 0.9503 | 0.2462 | 0.2419 |
| 60 | 10 | 0.9430 | 0.9518 | 0.2479 | 0.2635 |
| 100 | 10 | 0.9190 | 0.9524 | 0.2715 | 0.3095 |
| 250 | 10 | 0.9060 | 0.9631 | 0.2173 | 0.2887 |

interval for parameter $\theta_{jk}^*$ is given by

$$I_{jk}(\alpha) = \left[\widehat{\theta}_{jk}^u - \frac{\widehat{\xi}_{jk}}{\sqrt{n}}\Phi^{-1}\left(1-\frac{\alpha}{2}\right),\ \widehat{\theta}_{jk}^u + \frac{\widehat{\xi}_{jk}}{\sqrt{n}}\Phi^{-1}\left(1-\frac{\alpha}{2}\right)\right],$$

where $\widehat{\Theta}^u$ is the de-biased estimator and $\widehat{\xi}_{jk}$ is specified in (4.2). Throughout this section, we set $\alpha = 0.05$. For every parameter $\theta_{jk}^*$, we estimate the probability that the true value $\theta_{jk}^*$ is covered by the confidence interval $I_{jk}(\alpha)$ using its empirical version, i.e., $\widehat{\alpha}_{jk}$ is the percentage of times that $\theta_{jk}^*$ is covered by $I_{jk}(\alpha)$ in 100 replications. Next for $S = \text{supp}(\Omega_{12}^*)$, we define the average coverage probability over $S$ and over $S^c$ to be

$$\text{Avgcov}_S = \frac{1}{|S|}\sum_{(j,k)\in S}\widehat{\alpha}_{jk}, \quad \text{Avgcov}_{S^c} = \frac{1}{|S^c|}\sum_{(j,k)\in S^c}\widehat{\alpha}_{jk}. \quad (5.2)$$

We also calculate the average length of the confidence intervals over $S$ and $S^c$ and denote them as Avglen$_S$ and Avglen$_{S^c}$ respectively. The result of these four quantities over 100 replications can be seen in Table 4. The coverage probabilities over the support $S$ and the non-support $S^c$ are around 95%, which matches the significance level $\alpha = 0.05$. And the coverage probability over $S$ decreases as the dimension $d$ increases, as expected.

In Figure 3, we show the QQ-plot of $\sqrt{n}\cdot(\widehat{\theta}_{jk}^u - \theta_{jk}^*)/\widehat{\xi}_{jk}$ when $d = 250$. We choose $(j,k) \in \{(126,1), (126,2), (126,3)\}$ to present. As we can see, the scattered points of $\sqrt{n}\cdot(\widehat{\theta}_{jk}^u - \theta_{jk}^*)/\widehat{\xi}_{jk}$ in 100 replications are close to the line with zero intercept and unit slope.

## 5.2 fMRI Data

In this section, we apply our estimation and inference methods to an fMRI data studied in (Chen et al., 2016). This data set includes fMRI measurements of 17 subjects while they were watching a 23-minute movie (BBC's "Sherlock"). The fMRI measurements were made every 1.5 seconds, thus in total we have 945 brain images for each subject. As described in Chen et al. (2016), the 23-minute movie is divided into 26 scenes for further analysis. For the original fMRI data, there are 271,633 voxels measured. We adopt the method introduced in Baldassano et al. (2015) to reduce the dimension to 172 regions of interest (ROIs). We use the average of the first eight subjects as $X_{\mathcal{G}_1}$ and the average of the remaining nine subjects as $X_{\mathcal{G}_2}$ for conducting Inter-Subject Analysis. For preprocessing, each ROI is

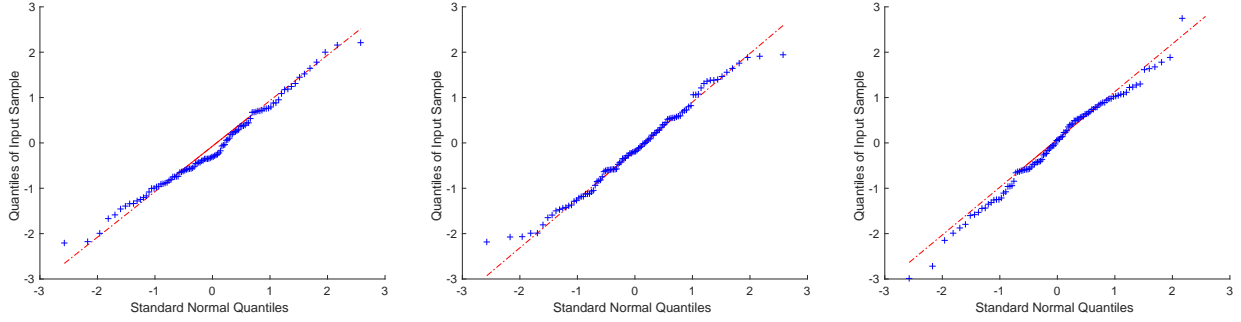

(a) $\sqrt{n} \cdot (\widehat{\theta}^u_{126,1} - \theta^*_{126,1})/\widehat{\xi}_{126,1}$  (b) $\sqrt{n} \cdot (\widehat{\theta}^u_{126,2} - \theta^*_{126,2})/\widehat{\xi}_{126,2}$  (c) $\sqrt{n} \cdot (\widehat{\theta}^u_{126,3} - \theta^*_{126,3})/\widehat{\xi}_{126,3}$

Figure 3: The QQ-plot of $\sqrt{n} \cdot (\widehat{\theta}^u_{jk} - \theta^*_{jk})/\widehat{\xi}_{jk}$ for $(j,k) = (126,1), (126,2)$ and $(126,3)$.

standardized to have zero mean and unit variance. For estimation, the tuning parameter $\lambda$ is chosen through cross validation. In inference, we threshold the de-biased estimator at level $\Phi^{-1}(1 - 4\alpha/d^2) \cdot \widehat{\xi}_{jk}/\sqrt{n}$, where $\alpha = 0.05$ and $4/d^2$ accounts for the Bonferroni correction in multiple hypothesis testing. We pick the eighth scene and the fifteenth scene for presentation. Scene 8 represents a press conference held by the police department to describe the recent suicides. Scene 15 contains the first meeting of Sherlock and Watson during which Sherlock shows his deduction talent to Watson.

In Figure 4, we show the brain networks for these two scenes estimated by our method. Each purple circle represents a region of interest (ROI). We also show the snapshots of both the left and the right brain hemispheres in Figure 5. The color represents the degree of the ROIs in the inter-subject conditional independence graph. And a redder area corresponds to the ROI with higher degree. As we can see, for the eighth scene when the press conference took place, the visual cortex and auditory cortex are highly activated since the subjects were mostly receiving audio and visual information from the press conference. The high activation of the visual and auditory cortices are ubiquitous in all 26 scenes. This makes sense since the subjects were under an audio-visual stimulus ("BBC's Sherlock"). This also matches the results in Chen et al. (2016). More specifically, during the fifteenth scene when Sherlock and Watson met, we can see that the prefrontal cortex especially the dorsolateral prefrontal cortex (DL-PFC) has a large degree. DL-PFC is known for its function in working memory and abstract reasoning (Miller and Cummings, 2007). And this coincides with scene 15 since the subjects might reason about Sherlock's deduction about Watson's job.

# 6 Extensions

In this section, we introduce two extensions to our methods and theories. Section 6.1 is devoted to extensions to non-Gaussian distributions and in Section 6.2, we generalize the Inter-Subject Analysis (ISA) to multiple subjects.

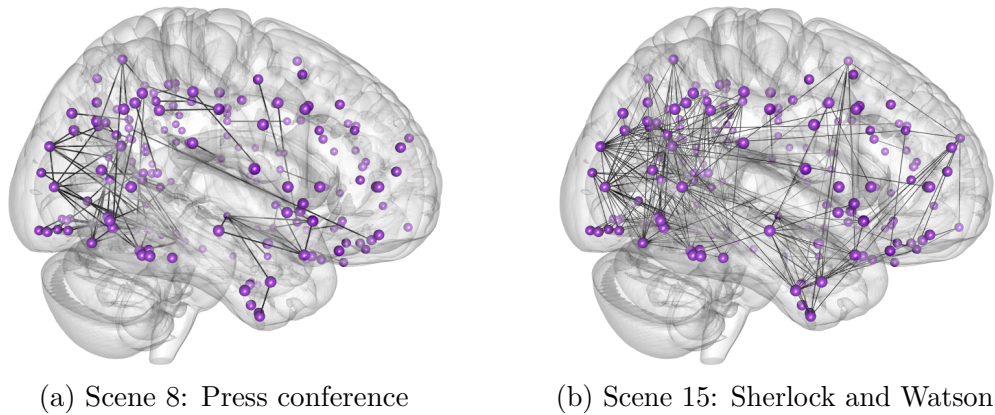

(a) Scene 8: Press conference  (b) Scene 15: Sherlock and Watson

Figure 4: The brain networks for two difference scenes. Each purple circle represents an ROI and the black edges represent the graph edges.

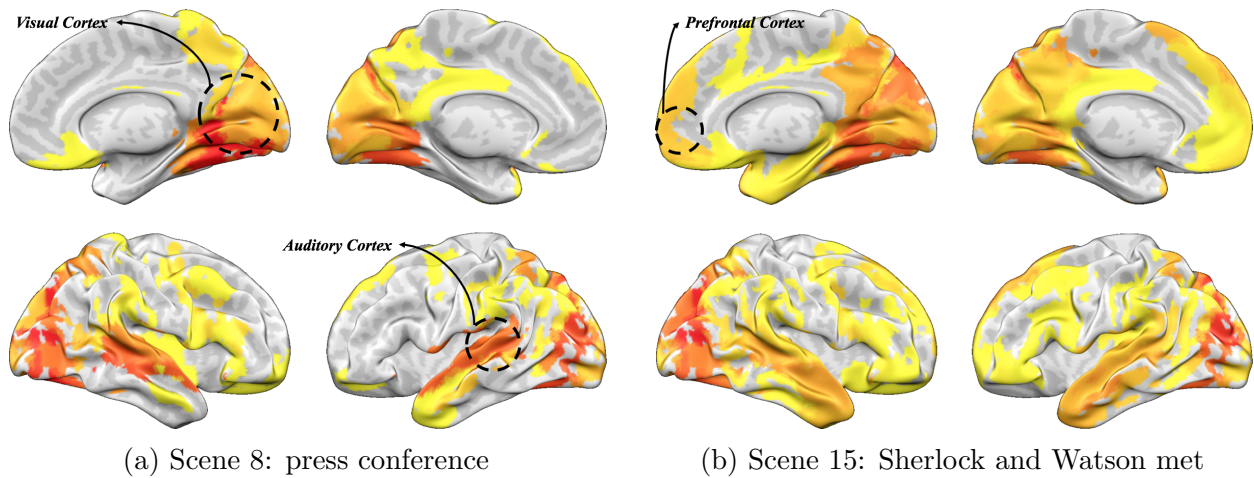

(a) Scene 8: press conference  (b) Scene 15: Sherlock and Watson met

Figure 5: the brain images for both left and right hemisphere.

## 6.1 Non-Gaussian Distributions

In this section, we extend our estimation methods to non-Gaussian distributions including the nonparanormal distribution studied in Liu et al. (2012) and Xue et al. (2012) and mixed data distributions studied in Fan et al. (2016). First, we introduce the definition of the nonparanormal distribution.

**Definition 6.1.** (Nonparanormal Distribution) Let $f = \{f_1, \ldots, f_d\}$ be a set of monotone univariate functions and let $\Sigma^* \in \mathbb{R}^{d \times d}$ be a positive definite correlation matrix with the diagonal being all 1's. A $d$-dimensional random vector $X = (X_1, \ldots, X_d)^\top$ follows a nonparanormal distribution $X \sim \text{NPN}_d(f, \Sigma^*)$ if $f(X) = (f_1(X_1), \ldots, f_d(X_d))^\top \sim N(0, \Sigma^*)$.

A close look at the proofs of Theorem 4.1 will show that the validity of the STRINGS estimator only depends on the $\ell_{\max}$-norm of the difference between $\widehat{\Sigma}$ and $\Sigma^*$. If we can obtain an estimator $\widehat{\Sigma}$ which satisfies $\|\widehat{\Sigma} - \Sigma^*\|_{\max} = \mathcal{O}_{\mathbb{P}}(\sqrt{\log d/n})$, then our estimation procedure is still valid. And this is the case for the nonparanormal distribution. The estimator $\widehat{\Sigma}$ for the nonparanormal distribution comes from the Kendall's tau statistics. Let $(j, k) \in [d] \times [d]$, define the $(j, k)$-th Kendall's tau statistic as

$$\widehat{\tau}_{jk} = \frac{2}{n(n-1)} \sum_{1 \leq i < i' \leq n} \text{sign}[(X_j^i - X_j^{i'})(X_k^i - X_k^{i'})].$$

$\widehat{\tau}_{jk}$ can be viewed as the nonparametric correlation between $X_j$ and $X_k$. And we define the following estimator $\widehat{\Sigma}^\tau = [\widehat{\Sigma}_{jk}^\tau]$ for the unknown correlation matrix $\Sigma^*$:

$$\widehat{\Sigma}_{jk}^\tau = \begin{cases} \sin(\frac{\pi}{2}\widehat{\tau}_{jk}), & \text{if } j \neq k, \\ 1, & \text{if } j = k. \end{cases}$$

By Theorem 4.2 in Liu et al. (2012), we have $\|\widehat{\Sigma}^\tau - \Sigma^*\|_{\max} \leq 2.45\pi\sqrt{\log d/n}$ with high probability. Based on this, we could get the following theorem.

**Theorem 6.2.** (Inter-Subject Analysis for Nonparanormal Distributions) Suppose the assumptions in Theorem 4.1 are satisfied. Then there exists a constant $C > 0$ such that for sufficiently large $n$, if $\lambda = 2C\sqrt{\log d/n}$, with probability $1 - 4d^{-1}$, we have

$$\|\widehat{\Theta} - \Theta^*\|_F \leq \frac{14C}{\rho_{\min}^2}\sqrt{\frac{s^* \log d}{n}} \quad \text{and} \quad \|\widehat{\Theta} - \Theta^*\|_{1,1} \leq \frac{56C}{\rho_{\min}^2}s^*\sqrt{\frac{\log d}{n}}.$$

We can see that Theorem 6.2 shares the same statistical rate with Theorem 4.1 under the Gaussian distribution. In the following we introduce the definition of the latent Gaussian copula model for binary data described in Fan et al. (2016).

**Definition 6.3.** (Binary Data) Let $X = (X_1, \ldots, X_d)^\top \in \{0, 1\}^d$ be a $d$-dimensional 0/1 random vector. We say $X$ satisfies latent Gaussian copula model if there exists a $d$-dimensional random vector $Z = (Z_1, \ldots, Z_d)^\top \sim \text{NPN}_d(f, \Sigma^*)$ such that

$$X_j = \mathbb{1}_{Z_j > C_j} \quad \text{for all } j = 1, \ldots, d,$$

where $C = (C_1, \ldots, C_d)$ is a vector of constants. Then we denote $X \sim \text{LNPN}(f, \Sigma^*, C)$.

The sample estimator of $\Sigma^*$ is also built upon the Kendall's tau statistics. However it is more involved than in the nonparanormal distribution. We omit the details and denote the sample estimator to be $\widehat{\Sigma}^b$. Then by Theorem 3.1 in Fan et al. (2016), we also have $\|\widehat{\Sigma}^b - \Sigma^*\|_{\max} = \mathcal{O}_{\mathbb{P}}(\sqrt{\log d/n})$. This will give us similar results as in Theorem 6.2.

In all, we can see that with a suitable choice of the sample covariance estimator $\widehat{\Sigma}$, our STRINGS estimator can be easily adapted to non-Gaussian distributions.

## 6.2 ISA with Multiple Subjects

In this section, we discuss the extension of ISA to multiple subjects.

Let $X = (X_1, \ldots, X_d)^\top$ be a $d$-dimensional random vector. Let $\mathcal{G}_1 \ldots, \mathcal{G}_L$ be $L$ disjoint subsets of $\{1, \ldots, d\}$ with cardinality $|\mathcal{G}_\ell| = d_\ell$ and $\sum_{\ell=1}^L d_\ell = d$. $X_{\mathcal{G}_\ell}$ represents features of the $\ell$-th subject. Let $\Sigma^* \in \mathbb{R}^{d \times d}$ be the covariance matrix of $X$, with $\Sigma^*_{jk} \in \mathbb{R}^{d_j \times d_k}$ being the covariance between $X_{\mathcal{G}_j}$ and $X_{\mathcal{G}_k}$. For the precision matrix $\Omega^* = (\Sigma^*)^{-1}$, we use $\Omega^*_{jk}$ to denote the dependency between $X_{\mathcal{G}_j}$ and $X_{\mathcal{G}_k}$. Define $\Sigma^*_\mathcal{G} = \text{diag}(\Sigma^*_{11}, \ldots, \Sigma^*_{LL})$ and $\Theta^* = \Omega^* - (\Sigma^*_\mathcal{G})^{-1}$, we have a similar observation.

**Proposition 6.4.** *If $L = \mathcal{O}(1)$ and $\|\Omega^*_{jk}\|_0 \leq s$ for all $j \neq k$, then $\|\Theta^*\|_0 = \mathcal{O}(s^2)$.*

Given Proposition 6.4, we can see that STRINGS estimator and the "untangle and chord" procedure are also valid for multiple subjects analysis.

# References

Ashby, F. G. (2011). *Statistical analysis of fMRI data.* MIT press.

Baldassano, C., Beck, D. M. and Fei-Fei, L. (2015). Parcellating connectivity in spatial maps. *PeerJ* **3** e784.

Banerjee, O., Ghaoui, L. E. and dAspremont, A. (2008). Model selection through sparse maximum likelihood estimation for multivariate gaussian or binary data. *Journal of Machine Learning Research* **9** 485–516.

Bartels, A. and Zeki, S. (2004). Functional brain mapping during free viewing of natural scenes. *Human brain mapping* **21** 75–85.

Belitski, A., Gretton, A., Magri, C., Murayama, Y., Montemurro, M. A., Logothetis, N. K. and Panzeri, S. (2008). Low-frequency local field potentials and spikes in primary visual cortex convey independent visual information. *The Journal of Neuroscience* **28** 5696–5709.

Bickel, P. J. and Levina, E. (2008). Regularized estimation of large covariance matrices. *The Annals of Statistics* 199–227.

Cai, T., Liu, W. and Luo, X. (2011). A constrained 1 minimization approach to sparse precision matrix estimation. *Journal of the American Statistical Association* **106** 594–607.

Cai, T. T., Liu, W., Zhou, H. H. et al. (2016). Estimating sparse precision matrix: Optimal rates of convergence and adaptive estimation. *The Annals of Statistics* **44** 455–488.

Chen, J., Leong, Y. C., Norman, K. A. and Hasson, U. (2016). Shared experience, shared memory: a common structure for brain activity during naturalistic recall. *bioRxiv* 035931.

Fan, J., Liu, H., Ning, Y. and Zou, H. (2016). High dimensional semiparametric latent graphical model for mixed data. *Journal of the Royal Statistical Society: Series B* .

Friedman, J., Hastie, T. and Tibshirani, R. (2008). Sparse inverse covariance estimation with the graphical lasso. *Biostatistics* **9** 432–441.

Gu, Q., Cao, Y., Ning, Y. and Liu, H. (2015). Local and global inference for high dimensional Gaussian copula graphical models. *arXiv preprint arXiv:1502.02347* .

Hartley, T., Maguire, E. A., Spiers, H. J. and Burgess, N. (2003). The well-worn route and the path less traveled: Distinct neural bases of route following and wayfinding in humans. *Neuron* **37** 877 – 888.

Hasson, U., Nir, Y., Levy, I., Fuhrmann, G. and Malach, R. (2004). Intersubject synchronization of cortical activity during natural vision. *science* **303** 1634–1640.

Horwitz, B. and Rapoport, S. I. (1988). Partial correlation coefficients approximate. *J Nucl Med* **29** 392–399.

Huang, S., Li, J., Sun, L., Liu, J., Wu, T., Chen, K., Fleisher, A., Reiman, E. and Ye, J. (2009). Learning brain connectivity of alzheimer's disease from neuroimaging data. In *Advances in Neural Information Processing Systems*.

Ipsen, I. C. (2009). *Numerical matrix analysis: Linear systems and least squares*. Siam.

Isserlis, L. (1916). On certain probable errors and correlation coefficients of multiple frequency distributions with skew regression. *Biometrika* **11** 185–190.

Jankova, J., van de Geer, S. et al. (2015). Confidence intervals for high-dimensional inverse covariance estimation. *Electronic Journal of Statistics* **9** 1205–1229.

Javanmard, A. and Montanari, A. (2014). Confidence intervals and hypothesis testing for high-dimensional regression. *Journal of Machine Learning Research* **15** 2869–2909.

Kolar, M., Liu, H. and Xing, E. P. (2014). Graph estimation from multi-attribute data. *Journal of Machine Learning Research* **15** 1713–1750.

Lee, H., Lee, D. S., Kang, H., Kim, B.-N. and Chung, M. K. (2011). Sparse brain network recovery under compressed sensing. *IEEE Transactions on Medical Imaging* **30** 1154–1165.

Liu, H., Han, F., Yuan, M., Lafferty, J. and Wasserman, L. (2012). High-dimensional semiparametric gaussian copula graphical models. *The Annals of Statistics* 2293–2326.

Liu, S., Suzuki, T., Sugiyama, M. and Fukumizu, K. (2015). Structure learning of partitioned markov networks. *arXiv preprint arXiv:1504.00624*.

Marrelec, G., Krainik, A., Duffau, H., Pélégrini-Issac, M., Lehéricy, S., Doyon, J. and Benali, H. (2006). Partial correlation for functional brain interactivity investigation in functional mri. *Neuroimage* **32** 228–237.

Mechler, F., Victor, J. D., Purpura, K. P. and Shapley, R. (1998). Robust temporal coding of contrast by v1 neurons for transient but not for steady-state stimuli. *Journal of Neuroscience* **18** 6583–6598.

Meinshausen, N. and Bühlmann, P. (2006). High-dimensional graphs and variable selection with the lasso. *The annals of statistics* 1436–1462.

Miller, B. L. and Cummings, J. L. (2007). *The human frontal lobes: Functions and disorders*. Guilford press.

Negahban, S. and Wainwright, M. J. (2011). Estimation of (near) low-rank matrices with noise and high-dimensional scaling. *The Annals of Statistics* 1069–1097.

Ning, Y. and Liu, H. (2014). A general theory of hypothesis tests and confidence regions for sparse high dimensional models. *arXiv preprint arXiv:1412.8765*.

Petersen, K. B. and Pedersen, M. S. (2012). The matrix cookbook. Version 20121115.

Ravikumar, P., Wainwright, M. J., Raskutti, G., Yu, B. et al. (2011). High-dimensional covariance estimation by minimizing 1-penalized log-determinant divergence. *Electronic Journal of Statistics* **5** 935–980.

Rohde, A., Tsybakov, A. B. et al. (2011). Estimation of high-dimensional low-rank matrices. *The Annals of Statistics* **39** 887–930.

Rothman, A. J., Bickel, P. J., Levina, E., Zhu, J. et al. (2008). Sparse permutation invariant covariance estimation. *Electronic Journal of Statistics* **2** 494–515.

Simony, E., Honey, C. J., Chen, J., Lositsky, O., Yeshurun, Y., Wiesel, A. and Hasson, U. (2016). Dynamic reconfiguration of the default mode network during narrative comprehension. *Nature Communications* **7**.

Van de Geer, S., Bühlmann, P., Ritov, Y., Dezeure, R. et al. (2014). On asymptotically optimal confidence regions and tests for high-dimensional models. *The Annals of Statistics* **42** 1166–1202.

Varoquaux, G., Gramfort, A., Poline, J. B. and Thirion, B. (2012). Markov models for fmri correlation structure: is brain functional connectivity small world, or decomposable into networks? *Journal of Physiology-Paris* **106** 212–221.

Vershynin, R. (2010). Introduction to the non-asymptotic analysis of random matrices. *arXiv preprint arXiv:1011.3027*.

Xia, Y., Cai, T. and Cai, T. T. (2016). Multiple testing of submatrices of a precision matrix with applications to identification of between pathway interactions. *J. Am. Stat. Assoc.* to appear.

Xue, L., Zou, H. et al. (2012). Regularized rank-based estimation of high-dimensional nonparanormal graphical models. *The Annals of Statistics* **40** 2541–2571.

Yao, H., Shi, L., Han, F., Gao, H. and Dan, Y. (2007). Rapid learning in cortical coding of visual scenes. *Nature neuroscience* **10** 772–778.

Yuan, M. (2010). High dimensional inverse covariance matrix estimation via linear programming. *Journal of Machine Learning Research* **11** 2261–2286.

Yuan, M. and Lin, Y. (2007). Model selection and estimation in the gaussian graphical model. *Biometrika* **94** 19–35.

Yuan, X.-T. and Zhang, T. (2014). Partial gaussian graphical model estimation. *IEEE Transactions on Information Theory* **60** 1673–1687.

Zacks, J. M., Braver, T. S., Sheridan, M. A., Donaldson, D. I., Snyder, A. Z., Ollinger, J. M., Buckner, R. L. and Raichle, M. E. (2001). Human brain activity time-locked to perceptual event boundaries. *Nature neuroscience* **4** 651–655.

Zhang, C.-H. and Zhang, S. S. (2014). Confidence intervals for low dimensional parameters in high dimensional linear models. *Journal of the Royal Statistical Society: Series B (Statistical Methodology)* **76** 217–242.

Zhao, S. D., Cai, T. T. and Li, H. (2014). Direct estimation of differential networks. *Biometrika* **101** 253–268.

# A Proof of the Key Observation

In this section, we give the proof of the key observation in Section 1.

*Proof of the Key Observation.* By the definition of $\Theta^*$, we have

$$\Theta^* = \Omega^* - (\Sigma_{\mathcal{G}}^*)^{-1} = \begin{bmatrix} \Sigma_1^* & \Sigma_{12}^* \\ \Sigma_{12}^{*\top} & \Sigma_2^* \end{bmatrix}^{-1} - \begin{bmatrix} \Sigma_1^* & 0 \\ 0 & \Sigma_2^* \end{bmatrix}^{-1} = \begin{bmatrix} \Theta_1^* & \Theta_{12}^* \\ \Theta_{12}^{*\top} & \Theta_2^* \end{bmatrix},$$

where $\Theta_1^*, \Theta_2^*$ and $\Theta_{12}^*$ are submatrices of $\Theta^*$. By the formula for matrix inversion in block form, we have $\Theta_1^* = (\Sigma_1^* - \Sigma_{12}^* \Sigma_2^{*-1} \Sigma_{12}^{*\top})^{-1} - \Sigma_1^{*-1}$, $\Theta_2^* = (\Sigma_2^* - \Sigma_{12}^{*\top} \Sigma_1^{*-1} \Sigma_{12}^*)^{-1} - \Sigma_2^{*-1}$ and $\Theta_{12}^* = \Omega_{12}^* = -\Sigma_1^{*-1} \Sigma_{12}^* (\Sigma_2^* - \Sigma_{12}^{*\top} \Sigma_1^{*-1} \Sigma_{12}^*)^{-1}$. Further by Woodbury's identity, we have

$$\Theta_1^* = \Sigma_1^{*-1} \Sigma_{12}^* (\Sigma_2^* - \Sigma_{12}^{*\top} \Sigma_1^{*-1} \Sigma_{12}^*)^{-1} \Sigma_{12}^{*\top} \Sigma_1^{*-1} = -\Omega_{12}^* \Sigma_{12}^{*\top} \Sigma_1^{*-1}. \tag{A.1}$$

And similarly we have the following identity for $\Theta_2^*$:

$$\Theta_2^* = \Sigma_2^{*-1} \Sigma_{12}^{*\top} (\Sigma_1^* - \Sigma_{12}^* \Sigma_2^{*-1} \Sigma_{12}^{*\top})^{-1} \Sigma_{12}^* \Sigma_2^{*-1} = -\Omega_{12}^{*\top} \Sigma_{12}^* \Sigma_2^{*-1}. \tag{A.2}$$

By the assumption that $\|\mathrm{vec}(\Omega_{12}^*)\|_0 \leq s$, we know that there are at most $s$ nonzero columns and $s$ nonzero rows in $\Omega_{12}^*$. By the relationship in (A.1), we can see that there are at most $s$ nonzero rows in $\Theta_1^*$. Moreover, since $\Theta_1^*$ is symmetric, we get that there are at most $s$ nonzero columns in $\Theta_1^*$. Hence $\|\mathrm{vec}(\Theta_1^*)\|_0 \leq s^2$. Applying the same argument to (A.2), we have $\|\mathrm{vec}(\Theta_2^*)\|_0 \leq s^2$. In all, we have $\|\mathrm{vec}(\Theta^*)\|_0 \leq 2s^2 + 2s$. □

# B Computational Algorithm

In this section, we derive the updates for optimizing (2.6) using alternating direction method of multipliers (ADMM).

We can rewrite (2.6) into the following form:

$$\widehat{\Theta} = \underset{W \in \mathbb{R}^{d \times d}}{\arg\min} \quad \mathrm{Tr}(W\widehat{\Sigma}) - \log|Y| + \lambda \|Z\|_{1,1} \tag{B.1}$$
$$\text{subject to} \quad W - Z = 0,$$
$$\widehat{\Sigma}_{\mathcal{G}} W \widehat{\Sigma}_{\mathcal{G}} + \widehat{\Sigma}_{\mathcal{G}} - Y = 0.$$

The augmented Lagrangian of (B.1) can be written as

$$L(W, Y, Z, U_1, U_2) = \mathrm{Tr}(W\widehat{\Sigma}) - \log|Y| + \lambda \|Z\|_{1,1} + \mathrm{Tr}[U_1(W - Z)]$$
$$+ \mathrm{Tr}[U_2(\widehat{\Sigma}_{\mathcal{G}} W \widehat{\Sigma}_{\mathcal{G}} + \widehat{\Sigma}_{\mathcal{G}} - Y)] + \frac{\rho}{2}\|W - Z\|_F^2 + \frac{\rho}{2}\|\widehat{\Sigma}_{\mathcal{G}} W \widehat{\Sigma}_{\mathcal{G}} + \widehat{\Sigma}_{\mathcal{G}} - Y\|_F^2.$$

ADMM is an iterative method that alternatively optimize over the primal variables $W, Y, Z$ and dual variables $U_1, U_2$. In the $(k+1)$-th step, for $W$, we solve the following:

$$\begin{aligned} W^{k+1} &= \arg\min_W L(X, Y^k, Z^k, U_1^k, U_2^k) \\ &= \arg\min_W \text{Tr}(X\widehat{\Sigma}) + \text{Tr}[U_1^k(W - Z^k)] + \text{Tr}[U_2^k(\widehat{\Sigma}_\mathcal{G} W \widehat{\Sigma}_\mathcal{G} + \widehat{\Sigma}_\mathcal{G} - Y^k)] \\ &\quad + \frac{\rho}{2}\|W - Z^k\|_F^2 + \frac{\rho}{2}\|\widehat{\Sigma}_\mathcal{G} W \widehat{\Sigma}_\mathcal{G} + \widehat{\Sigma}_\mathcal{G} - Y^k\|_F^2. \end{aligned} \quad (B.2)$$

The objective is a quadratic function of $W$ and the optimality condition of (B.2) is given by

$$\widehat{\Sigma} + U_1^k + \widehat{\Sigma}_\mathcal{G} U_2^k \widehat{\Sigma}_\mathcal{G} + \rho(W^{k+1} - Z^k) + \rho(\widehat{\Sigma}_\mathcal{G})^2 W^{k+1}(\widehat{\Sigma}_\mathcal{G})^2 + \rho\widehat{\Sigma}_\mathcal{G}(\widehat{\Sigma}_\mathcal{G} - Y^k)\widehat{\Sigma}_\mathcal{G} = 0,$$

which is equivalent to the following simplified form

$$W^{k+1} + AW^{k+1}A = B^k, \quad (B.3)$$

where $A = (\widehat{\Sigma}_\mathcal{G})^2$ and $B^k = Z^k - \widehat{\Sigma}_\mathcal{G}(\widehat{\Sigma}_\mathcal{G} - Y^k)\widehat{\Sigma}_\mathcal{G} - 1/\rho(\widehat{\Sigma} + U_1^k + \widehat{\Sigma}_\mathcal{G} U_2^k \widehat{\Sigma}_\mathcal{G})$.

Let $A = VDV^\top$ be the eigen-decomposition of $A$, then multiplying (B.3) by $V^\top$ from left and $V$ from right would give $V^\top W^{k+1}V + DV^\top W^{k+1}VD = V^\top B^k V$. By viewing $V^\top W^k V$ as a new variable $T^k$, we have $T^k + DT^k D = V^\top B^k V$. Since $D$ is a diagonal matrix, $T^k$ can be easily solved as $T^k = (V^\top B^k V)./[\mathbf{1}\mathbf{1}^\top + \text{diag}(D)\text{diag}(D)]$, where ./ denotes the element-wise division. Then $W^{k+1}$ can be recovered by $W^{k+1} = VT^k V^\top$.

After obtaining $W^{k+1}$, we would solve the following optimization problem for $Y$:

$$\begin{aligned} Y^{k+1} &= \arg\min_Y L(W^{k+1}, Y, Z^k, U_1^k, U_2^k) \\ &= \arg\min_Y -\log|Y| + \text{Tr}[U_2^k(\widehat{\Sigma}_\mathcal{G} W^{k+1} \widehat{\Sigma}_\mathcal{G} + \widehat{\Sigma}_\mathcal{G} - Y)] + \frac{\rho}{2}\|\widehat{\Sigma}_\mathcal{G} W^{k+1} \widehat{\Sigma}_\mathcal{G} + \widehat{\Sigma}_\mathcal{G} - Y\|_F^2. \end{aligned} \quad (B.4)$$

Although the objective in (B.4) involves the log-determinant term, the minimizer enjoys a closed-form solution. The first order optimality condition of (B.4) is given by

$$-Y^{-1} - U_2^k + \rho[Y - (\widehat{\Sigma}_\mathcal{G} W^{k+1} \widehat{\Sigma}_\mathcal{G} + \widehat{\Sigma}_\mathcal{G})] = 0.$$

Define $C^k = U_2^k + \rho(\widehat{\Sigma}_\mathcal{G} W^{k+1} \widehat{\Sigma}_\mathcal{G} + \widehat{\Sigma}_\mathcal{G})$ with the eigen-decomposition $C^k = Q^k \Lambda^k Q^{k\top}$, then construct a diagonal matix $\widetilde{Y}^k$ with $\widetilde{Y}_{jj}^k = \left(\Lambda_{jj}^k + \sqrt{(\Lambda_{jj}^k)^2 + 4\rho}\right)/(2\rho)$, and the solution to (B.4) has the form $Y^{k+1} = Q^k \widetilde{Y}^k Q^{k\top}$.

The following updates for $Z, U_1$ and $U_2$ are easy, thus we omit here.

## C Proof of the rate of convergence for STRINGS

Here we outline the proof of Theorem 4.1 on the statistical rate of the STRINGS estimator. To simplify notations, let $S = \text{supp}(\Theta^*)$ be the support of $\Theta^*$. By Assumption (**A1**), we

**Algorithm 2** STRINGS: **S**parse edge es**T**imator for **I**ntense **N**uIS**A**nce **G**raph**S**

---

**Input**: $\mathbb{X} \in \mathbb{R}^{n \times d}$, where rows of $\mathbb{X}$ represent i.i.d samples from $N(0, \Sigma^*)$.
**Output**: $\widehat{\Theta}$, the STRINGS estimator for $\Theta^*$.
**Covariance Estimation**: Obtain the sample covariance $\widehat{\Sigma} = (1/n) \cdot \mathbb{X}^\top \mathbb{X}$ and $\widehat{\Sigma}_\mathcal{G} = \mathrm{diag}(\widehat{\Sigma}_1, \widehat{\Sigma}_2)$, where $\widehat{\Sigma}_1$ and $\widehat{\Sigma}_2$ are the diagonal submatrices of $\widehat{\Sigma}$.
**Preconditioning**: if $\lambda_{\min}(\widehat{\Sigma}_\mathcal{G}) = 0$, let $\widehat{\Sigma}_\mathcal{G} = \widehat{\Sigma}_\mathcal{G} + \sqrt{\log d/n} \cdot I$.
**Initialization**: Set $W^0 = Z^0 = 0$, $Y^0 = \widehat{\Sigma}_\mathcal{G}$ and $U_1^0 = U_2^0 = 0$. Fix the step size $\rho > 0$.
**Preprocessing**: Form $A = (\widehat{\Sigma}_\mathcal{G})^2$ and compute the eigen-decomposition of $A = VDV^\top$.
**for** $k = 0, 1, 2, \ldots$ **do**
  **W update**: Form $B^k = Z^k - \widehat{\Sigma}_\mathcal{G}(\widehat{\Sigma}_\mathcal{G} - Y^k)\widehat{\Sigma}_\mathcal{G} - (1/\rho) \cdot (\widehat{\Sigma} + U_1^k + \widehat{\Sigma}_\mathcal{G} U_2^k \widehat{\Sigma}_\mathcal{G})$. Let $T^k = (U^\top B^k U)./[\mathbf{1}\mathbf{1}^\top + \mathrm{diag}(D)\mathrm{diag}(D)^\top]$. Then set $W^{k+1} = VT^k V^\top$.
  **Y update**: Form $C^k = U_2^k + \rho(\widehat{\Sigma}_\mathcal{G} W^{k+1} \widehat{\Sigma}_\mathcal{G} + \widehat{\Sigma}_\mathcal{G})$ and its eigen-decomposition $C^k = Q^k \Lambda^k Q^{k\top}$. Then set $Y^{k+1} = Q^k \widetilde{Y}^k Q^{k\top}$, where $\widetilde{Y}^k$ is a diagonal matrix with $\widetilde{Y}_{jj}^k = (\Lambda_{jj}^k + \sqrt{(\Lambda_{jj}^k)^2 + 4\rho})/(2\rho)$.
  **Z update**: Set $Z^{k+1} = \tau_{\lambda/\rho}(W^{k+1} + U_1^k/\rho)$, where $\tau_a(v) = (v - a)_+ - (-v - a)_+$.
  **$U_1$ update**: Set $U_1^{k+1} = U_1^k + \rho(W^{k+1} - Z^{k+1})$.
  **$U_2$ update**: Set $U_2^{k+1} = U_2^k + \rho(\widehat{\Sigma}_\mathcal{G} W^{k+1} \widehat{\Sigma}_\mathcal{G} + \widehat{\Sigma}_\mathcal{G} - Y^{k+1})$.
**end for**
**return** $\widehat{\Theta} = W^k$.

---

have $|S| \leq s^*$. Define a cone $\mathbb{C} = \{\Delta \in \mathbb{R}^{d \times d} \mid \|\Delta_{S^c}\|_{1,1} \leq 3\|\Delta_S\|_{1,1}\}$. Further, define a set $\mathbb{C}^{(t)} = \mathbb{C} \cap \{\Delta \in \mathbb{R}^{d \times d} \mid \|\Delta\|_F = t\}$. Define a function $\mathcal{H}(\Delta) : \mathbb{R}^{d \times d} \to \mathbb{R}$ to be $\mathcal{H}(\Delta) = \mathcal{L}_n(\Theta^* + \Delta) + \lambda\|\Theta^* + \Delta\|_{1,1} - [\mathcal{L}_n(\Theta^*) + \lambda\|\Theta^*\|_{1,1}]$. Denote $\widehat{\Delta} = \widehat{\Theta} - \Theta^*$ and $t^* = (14C/\rho_{\min}^2)\sqrt{s^* \log d/n}$, which is the desired rate of convergence for $\|\widehat{\Delta}\|_F$ and $C$ is a constant specified in Lemma C.1.

We first state several technical lemmas of which the proofs are deferred to Section E. The first lemma bounds $\|\nabla \mathcal{L}_n(\Theta^*)\|_{\max}$, therefore connecting $\|\nabla \mathcal{L}_n(\Theta^*)\|_{\max}$ with $\lambda$.

**Lemma C.1.** ($\ell_{\max}$-norm Bound) Under Assumptions (**A1**)-(**A4**), then with probability at least $1 - 2d^{-1}$, we have $\|\nabla \mathcal{L}_n(\Theta^*)\|_{\max} \leq C\sqrt{\log d/n}$, where $C$ is a constant.

The following lemma provides the restricted strong convexity condition for the empirical loss function $\mathcal{L}_n(\Theta)$ in a neighborhood of $\Theta^*$.

**Lemma C.2.** (Restricted Strong Convexity) Under Assumptions (**A1**)-(**A4**), we have with probability at least $1 - 2d^{-1}$, for all $\Delta \in \mathbb{C}^{(t^*)}$ and $\mu \in [0, 1]$,

$$|\mathrm{vec}(\Delta)^\top \nabla^2 \mathcal{L}_n(\Theta^* + \mu\Delta)\mathrm{vec}(\Delta)| \geq \left(\rho_{\min}^2 - C' s^{*2} \sqrt{\frac{\log d}{n}}\right) \|\Delta\|_F^2,$$

where $C'$ is a constant. Moreover if $s^{*2}\sqrt{\log d/n} = o(1)$, for all $\Delta \in \mathbb{C}^{(t^*)}$ and $\mu \in [0, 1]$, we have $|\mathrm{vec}(\Delta)^\top \nabla^2 \mathcal{L}_n(\Theta^* + \mu\Delta)\mathrm{vec}(\Delta)| \geq (1/2)\rho_{\min}^2 \|\Delta\|_F^2$.

We are now ready to prove Theorem 4.1.

*Proof of Theorem 4.1.* Define the following two events

$$\mathcal{E}_1 = \left\{\|\nabla\mathcal{L}_n(\Theta^*)\|_{\max} \leq C\sqrt{\frac{\log d}{n}}, \text{ where } C \text{ is the same constant as in Lemma C.1}\right\},$$

$$\mathcal{E}_2 = \left\{\text{For all } \Delta \in \mathbb{C}^{(t^*)}, \mu \in [0,1], |\text{vec}(\Delta)^\top \nabla^2 \mathcal{L}_n(\Theta^* + \mu\Delta)\text{vec}(\Delta)| \geq \frac{\rho_{\min}^2}{2}\|\Delta\|_F^2\right\}.$$

By Lemma C.1 and Lemma C.2, we have $\mathbb{P}(\mathcal{E}_1) \geq 1 - 2d^{-1}$ and $\mathbb{P}(\mathcal{E}_2) \geq 1 - 2d^{-1}$. Thus

$$\mathbb{P}(\mathcal{E}_1 \cap \mathcal{E}_2) = 1 - \mathbb{P}(\mathcal{E}_1^c \cup \mathcal{E}_2^c) \geq 1 - \mathbb{P}(\mathcal{E}_1^c) - \mathbb{P}(\mathcal{E}_2^c) \geq 1 - 4d^{-1},$$

where we use the union bound that $\mathbb{P}(\mathcal{E}_1^c \cup \mathcal{E}_2^c) \leq \mathbb{P}(\mathcal{E}_1^c) + \mathbb{P}(\mathcal{E}_2^c)$. In the rest of the proof, we are always conditioning on the event $\mathcal{E} = \mathcal{E}_1 \cap \mathcal{E}_2$.

Under $\mathcal{E}_1$, we have $\lambda = 2C\sqrt{\log d/n} \geq 2\|\nabla\mathcal{L}_n(\Theta^*)\|_{\max}$. We first specify the space where $\widehat{\Delta}$ lies in when $\lambda$ is chosen as this. By the definition of $\mathcal{H}(\Delta)$ and $\widehat{\Delta}$, we have

$$\mathcal{H}(\widehat{\Delta}) = \mathcal{L}_n(\Theta^* + \widehat{\Delta}) + \lambda\|\Theta^* + \widehat{\Delta}\|_{1,1} - [\mathcal{L}_n(\Theta^*) + \lambda\|\Theta^*\|_{1,1}]$$
$$= \underbrace{\mathcal{L}_n(\Theta^* + \widehat{\Delta}) - \mathcal{L}_n(\Theta^*)}_{I_1} - \underbrace{(\lambda\|\Theta^*\|_{1,1} - \lambda\|\Theta^* + \widehat{\Delta}\|_{1,1})}_{I_2}. \quad (C.1)$$

For the first term, since $\mathcal{L}_n(\Theta)$ is a convex function, we have $I_1 \geq \text{Tr}[\nabla\mathcal{L}_n(\Theta^*)\widehat{\Delta}]$. Further using generalized Cauchy-Schwarz inequality, we get

$$\big|\text{Tr}[\nabla\mathcal{L}_n(\Theta^*)\widehat{\Delta}]\big| \leq \|\nabla\mathcal{L}_n(\Theta^*)\|_{\max}\|\widehat{\Delta}\|_{1,1} \leq \frac{\lambda}{2}\|\widehat{\Delta}\|_{1,1},$$

where we also use the fact that $\lambda \geq 2\|\nabla\mathcal{L}_n(\Theta^*)\|_{\max}$. Hence we get

$$I_1 \geq -\big|\text{Tr}[\nabla\mathcal{L}_n(\Theta^*)\widehat{\Delta}]\big| \geq -\frac{\lambda}{2}\|\widehat{\Delta}\|_{1,1}. \quad (C.2)$$

For the second term, by decomposing the $\ell_{1,1}$-norm into $S$ and $S^c$, we have

$$I_2 = \lambda\|\Theta_S^*\|_{1,1} - \lambda(\|\Theta_S^* + \widehat{\Delta}_S\|_{1,1} + \|\widehat{\Delta}_{S^c}\|_{1,1}) \leq \lambda\|\widehat{\Delta}_S\|_{1,1} - \lambda\|\widehat{\Delta}_{S^c}\|_{1,1}, \quad (C.3)$$

where we use the triangle inequality. Combining (C.1), (C.2) and (C.3) together with the fact that $\mathcal{H}(\widehat{\Delta}) \leq 0$, we get $-(\lambda/2)\|\widehat{\Delta}\|_{1,1} - (\lambda\|\widehat{\Delta}_S\|_{1,1} - \lambda\|\widehat{\Delta}_{S^c}\|_{1,1}) \leq 0$. And this gives us $\|\widehat{\Delta}_{S^c}\|_{1,1} \leq 3\|\widehat{\Delta}_S\|_{1,1}$, i.e., $\widehat{\Delta} \in \mathbb{C}$.

We further claim that if $\mathcal{H}(\Delta) > 0$ for all $\Delta \in \mathbb{C}^{(t)}$, we must have $\|\widehat{\Delta}\|_F \leq t$. We prove this argument by contradiction. Suppose we have $\|\widehat{\Delta}\|_F > t$. Take $\mu = t/\|\widehat{\Delta}\|_F < 1$, then we have $\|\mu\widehat{\Delta}\|_F = t$ and $\mu\widehat{\Delta} \in \mathbb{C}^{(t)}$ since $\widehat{\Delta} \in \mathbb{C}$ and $\mathbb{C}$ is a cone. We also have

$$\mathcal{H}(\mu\widehat{\Delta}) = \mathcal{H}\big((1-\mu)0 + \mu\widehat{\Delta}\big) \leq (1-\mu)\mathcal{H}(0) + \mu\mathcal{H}(\widehat{\Delta}) \leq 0,$$

where in the first inequality we use the fact that $\mathcal{H}(\Delta)$ is a convex function of $\Delta$ and in the second inequality we use the fact that $\mathcal{H}(0) = 0$ and $\mathcal{H}(\widehat{\Delta}) \leq 0$. Hence we find a matrix $\Delta^* = \mu\widehat{\Delta} \in \mathbb{C}^{(t)}$ and $\mathcal{H}(\Delta^*) \leq 0$. This contradicts our assumption that $\mathcal{H}(\Delta) > 0$ for all $\Delta \in \mathbb{C}^{(t)}$. Thus we prove the argument.

Based on this, it suffices to show that $\mathcal{H}(\Delta) > 0$ for all $\Delta \in \mathbb{C}^{(t^*)}$. For $\Delta \in \mathbb{C}^{(t^*)}$, we have

$$\mathcal{H}(\Delta) = \underbrace{\mathcal{L}_n(\Theta^* + \Delta) - \mathcal{L}_n(\Theta^*)}_{I_3} - \underbrace{(\lambda\|\Theta^*\|_{1,1} - \lambda\|\Theta^* + \Delta\|_{1,1})}_{I_4}. \tag{C.4}$$

For the first term, by the mean-value theorem, we have

$$I_3 = \text{Tr}[\nabla\mathcal{L}_n(\Theta^*)\Delta] + \frac{1}{2}\text{vec}(\Delta)^\top \nabla^2 \mathcal{L}_n(\widetilde{\Theta})\text{vec}(\Delta),$$

where $\widetilde{\Theta} = \Theta^* + \mu\Delta$ for some $\mu \in [0,1]$. Since we are conditioning on $\mathcal{E}$, we have $(1/2)\text{vec}(\Delta)^\top \nabla^2 \mathcal{L}_n(\widetilde{\Theta})\text{vec}(\Delta) \geq (\rho_{\min}^2/4)\|\Delta\|_F^2$. Combining with (C.2), we have

$$I_3 \geq \frac{\rho_{\min}^2}{4}\|\Delta\|_F^2 - \frac{\lambda}{2}\|\Delta\|_{1,1} = \frac{\rho_{\min}^2}{4}\|\Delta\|_F^2 - \frac{\lambda}{2}\|\Delta_S\|_{1,1} - \frac{\lambda}{2}\|\Delta_{S^c}\|_{1,1}. \tag{C.5}$$

For the second term, we have the same decomposition as in (C.3):

$$I_4 = \lambda\|\Theta^*_S\|_{1,1} - \lambda(\|\Theta^*_S + \Delta_S\|_{1,1} + \|\Delta_{S^c}\|_{1,1}) \leq \lambda\|\Delta_S\|_{1,1} - \lambda\|\Delta_{S^c}\|_{1,1}. \tag{C.6}$$

Combing (C.4), (C.5) and (C.6), we have

$$\mathcal{H}(\Delta) \geq \frac{\rho_{\min}^2}{4}\|\Delta\|_F^2 - \frac{\lambda}{2}\|\Delta_S\|_{1,1} - \frac{\lambda}{2}\|\Delta_{S^c}\|_{1,1} - (\lambda\|\Delta_S\|_{1,1} - \lambda\|\Delta_{S^c}\|_{1,1})$$
$$\geq \frac{\rho_{\min}^2}{4}\|\Delta\|_F^2 - \frac{3}{2}\lambda\sqrt{s^*}\|\Delta\|_F,$$

where in the second inequality we use $\|\Delta_S\|_{1,1} \leq \sqrt{s^*}\|\Delta_S\|_F \leq \sqrt{s^*}\|\Delta\|_F$. Since $\Delta \in \mathbb{C}^{(t^*)}$, we have $\|\Delta\|_F = t^*$. Hence following some algebra, we get $\mathcal{H}(\Delta) > 0$. Thus we prove the rate for the Frobenius norm.

For the $\ell_{1,1}$-norm, we use the fact that $\widehat{\Delta} \in \mathbb{C}$. Thus we have $\|\widehat{\Delta}_{S^c}\|_{1,1} \leq 3\|\widehat{\Delta}_S\|_{1,1}$. And this leads to

$$\|\widehat{\Delta}\|_{1,1} \leq 4\|\widehat{\Delta}_S\|_{1,1} \leq 4\sqrt{s^*}\|\widehat{\Delta}_S\|_F \leq 4\sqrt{s^*}\|\widehat{\Delta}\|_F,$$

where in the second inequality we use the Holder's inequality and in the third we use the fact that $\|\widehat{\Delta}_S\|_F \leq \|\widehat{\Delta}\|_F$. Further because $\|\widehat{\Delta}\|_F \leq t^*$, we have $\|\widehat{\Delta}\|_{1,1} \leq 4\sqrt{s^*} \cdot t^*$. This gives us the desired rate for $\ell_{1,1}$ norm. $\square$

# D  Proof of the De-biased Estimator

In this section, we outline the proof of Theorem 4.2 and Lemma 4.4. We first provide some technical lemmas of which the proofs are deferred to Section F.

For asymptotics, we use the following standard notations: we write $f(n) = \mathcal{O}(g(n))$ if $f(n) \leq Cg(n)$ for some positive constant $C$ and all sufficiently large $n$. And $f(n) \asymp g(n)$ means $f(n) = \mathcal{O}(g(n))$ and $g(n) = \mathcal{O}(f(n))$.

The first lemma specifies the desired properties of the adjustment matrices $M$ and $P$ which will be useful to bound the entries in the Remainder term in (3.4).

**Lemma D.1.** (Bias Correction Matrices) Under Assumptions (**A1**) $-$ (**A5**), for $\lambda' = C'\sqrt{\log d/n}$, where $C'$ is a sufficiently large constant in (3.5) and (3.6), we have

$$\|M\|_\infty = \mathcal{O}_\mathbb{P}(1) \quad \text{and} \quad \|P\|_\infty = \mathcal{O}_\mathbb{P}(1),$$

$$\|M\widehat{\Sigma} - I\|_{\max} = \mathcal{O}_\mathbb{P}(\sqrt{\frac{\log d}{n}}) \quad \text{and} \quad \|P\widehat{\Sigma}_\mathcal{G} - I\|_{\max} = \mathcal{O}_\mathbb{P}(\sqrt{\frac{\log d}{n}}).$$

Given the good properties of the bias correction matrices $M$ and $P$, the next lemma bounds the Remainder term in (3.4).

**Lemma D.2.** (Remainder Term) Suppose the conditions in Theorem 4.2 hold. We have $\|\text{Remainder}\|_{\max} = \mathcal{O}_\mathbb{P}(s^* \log d/n)$.

The next lemma establishes the lower bound for the asymptotic variance in (4.1).

**Lemma D.3.** (Variance Lower Bound) Under Assumptions of Theorem 4.2, $1/\xi_{jk} = \mathcal{O}_\mathbb{P}(1)$.

The third lemma provides an explicit formula for the 4-th order moments of multivariate Gaussian distribution. It will be used to show the asymptotic variance of the Leading term.

**Lemma D.4.** (4-th Order Moments) For a random vector $X \sim N(0, \Sigma)$ and four deterministic matrices $A, B, C, D$ of appropriate sizes, we have

$$\mathbb{E}[(AX)(BX)^\top (CX)(DX)^\top] = (A\Sigma B^\top)(C\Sigma D^\top) + (A\Sigma C^\top)(B\Sigma D^\top) + \text{Tr}(B\Sigma C^\top)(A\Sigma D^\top).$$

*Proof.* This can be found in Petersen and Pedersen (2012). □

The fourth lemma characterizes the tail behavior of a certain random variable. It will later be used to show the asymptotic normality of the Leading term in (3.3).

**Lemma D.5.** (Tail Bound) Let $X \sim N(0, \Sigma^*)$. For any deterministic vectors $u, v$ with $v_{\mathcal{G}_1} = 0$, $\|u\|_1 = \mathcal{O}(1)$ and $\|v\|_1 = \mathcal{O}(1)$, define the following random variable

$$Z = -u^\top X X^\top (I + \Theta^* \Sigma_\mathcal{G}^*) v + u^\top (I - \Sigma^* \Theta^*) X_{\mathcal{G}_2} X_{\mathcal{G}_2}^\top v + u^\top \Sigma_\mathcal{G}^* v - u^\top \Sigma^* v.$$

We have that $Z$ is a sub-exponential random variable with $\|Z\|_{\psi_1} = \mathcal{O}(1)$.

Now we are ready to prove Theorem 4.2.

*Proof of Theorem 4.2.* From (3.3) and (3.4), we have for $j \in [d_1]$ and $k \in [d_1 + 1, d]$

$$\sqrt{n} \cdot (\widehat{\theta}_{jk}^u - \theta_{jk}^*) = \sqrt{n}\, \text{Leading}_{jk} + \sqrt{n}\, \text{Remainder}_{jk}. \tag{D.1}$$

By Lemma D.2, we have $\|\text{Remainder}\|_{\max} = \mathcal{O}_{\mathbb{P}}(s^* \log d / n)$. Under the scaling condition that $s^* \log d / \sqrt{n} = o(1)$, we have $\sqrt{n}\, \text{Remainder}_{jk} = o_{\mathbb{P}}(1)$. By (3.3), we have

$$\sqrt{n}\, \text{Leading}_{jk} = \frac{1}{\sqrt{n}} \sum_{i=1}^n \Big\{ M_{j*} \Sigma_{\mathcal{G}}^* P_{k*}^\top - M_{j*} \Sigma^* P_{k*}^\top - M_{j*} X_i X_i^\top (I + \Theta^* \Sigma_{\mathcal{G}}^*) P_{k*}^\top$$
$$+ M_{j*} (I - \Sigma^* \Theta^*)(X_{i,\mathcal{G}_1} X_{i,\mathcal{G}_1}^\top + X_{i,\mathcal{G}_2} X_{i,\mathcal{G}_2}^\top) P_{k*}^\top \Big\}, \tag{D.2}$$

where $X_i$ denotes the $i$-th sample and $X_{i,\mathcal{G}_1}$ and $X_{i,\mathcal{G}_2}$ are features of $X_i$ corresponding to $\mathcal{G}_1$ and $\mathcal{G}_2$. Further due to the block structure of $P$ and the fact that $k \in [d_1 + 1, d]$, we have for every $i \in [n]$, $X_{i,\mathcal{G}_1}^\top P_{k*}^\top = 0$. Thus (D.2) can be simplified as follows

$$\sqrt{n}\, \text{Leading}_{jk} = \frac{1}{\sqrt{n}} \sum_{i=1}^n \Big\{ M_{j*} \Sigma_{\mathcal{G}}^* P_{k*}^\top - M_{j*} \Sigma^* P_{k*}^\top - M_{j*} X_i X_i^\top (I + \Theta^* \Sigma_{\mathcal{G}}^*) P_{k*}^\top$$
$$+ M_{j*} (I - \Sigma^* \Theta^*) X_{i,\mathcal{G}_2} X_{i,\mathcal{G}_2}^\top P_{k*}^\top \Big\}. \tag{D.3}$$

To further simplify notations, for each $i \in [n]$, we define the following random variable

$$Z_{jk,i} = M_{j*} \Sigma_{\mathcal{G}}^* P_{k*}^\top - M_{j*} \Sigma^* P_{k*}^\top - M_{j*} X_i X_i^\top (I + \Theta^* \Sigma_{\mathcal{G}}^*) P_{k*}^\top + M_{j*} (I - \Sigma^* \Theta^*) X_{i,\mathcal{G}_2} X_{i,\mathcal{G}_2}^\top P_{k*}^\top.$$

Denote $S_n = \sum_{i=1}^n Z_{jk,i}$ and $s_n^2 = n \xi_{jk}^2$. Combining (D.1) and (D.3), we have

$$\sqrt{n} \cdot (\widehat{\theta}_{jk}^u - \theta_{jk}^*) / \xi_{jk} = S_n / s_n + o_{\mathbb{P}}(1) / \xi_{jk}, \tag{D.4}$$

where we also use the fact that $\sqrt{n}\, \text{Remainder}_{jk} = o_{\mathbb{P}}(1)$. By Lemma D.3, we have $o_{\mathbb{P}}(1) / \xi_{jk} = o_{\mathbb{P}}(1)$. Thus it remains to show that $S_n / s_n$ weakly converges to a standard normal distribution. We prove this using a conditioning argument.

For fixed $n$, $\mathcal{D}_1$ and $\mathcal{D}_2$ are two random variables (matrices) in $\mathbb{R}^{n \times d}$, where each row is a Gaussian random variable. Define the following set:

$$\Gamma = \{ \mathbb{X} \in \mathbb{R}^{n \times d} \,|\, \|\widehat{\Sigma}' - \Sigma^*\|_{\max} \leq C \sqrt{\frac{\log d}{n}} \},$$

where $\widehat{\Sigma}'$ is the sample covariance obtained from $\mathbb{X}$ and $C$ is a sufficiently large constant. We can see that $\Gamma$ is a set of data sets of which the sample covariance matrix is close to the population one.

Given any $\mathbb{X} \in \Gamma$, we have $M$ and $P$ are fixed since they purely depend on $\mathbb{X}$. By the proof of Lemma D.1, we can see that under $\mathbb{X} \in \Gamma$, $\|M\|_\infty = \mathcal{O}(1)$, $\|P\|_\infty = \mathcal{O}(1)$, $\|M\Sigma^* - I\|_{\max} = \mathcal{O}(\sqrt{\log d/n})$ and $\|P\Sigma_{\mathcal{G}}^* - I\|_{\max} = \mathcal{O}(\sqrt{\log d/n})$. Further, $Z_{jk,1}, \ldots, Z_{jk,n}$ are conditionally i.i.d random variables given $\mathcal{D}_2 = \mathbb{X}$. Its conditional mean is

$$\begin{aligned}\mathbb{E}[Z_{jk,i}|\mathcal{D}_2 = \mathbb{X}] &= M_{j*}\Sigma_{\mathcal{G}}^* P_{k*}^\top - M_{j*}\Sigma^* P_{k*}^\top - M_{j*}\Sigma^*(I + \Theta^*\Sigma_{\mathcal{G}}^*)P_{k*}^\top + M_{j*}^\top(I - \Sigma^*\Theta^*)\Sigma_{\mathcal{G}_2}^* P_{k*}^\top \\ &= M_{j*}\Sigma_{\mathcal{G}}^* P_{k*}^\top - M_{j*}\Sigma^* P_{k*}^\top - M_{j*}\Sigma_{\mathcal{G}}^* P_{k*}^\top + M_{j*}^\top(I - \Sigma^*\Theta^*)\Sigma_{\mathcal{G}}^* P_{k*}^\top \\ &= M_{j*}\Sigma_{\mathcal{G}}^* P_{k*}^\top - M_{j*}\Sigma^* P_{k*}^\top - M_{j*}\Sigma_{\mathcal{G}}^* P_{k*}^\top + M_{j*}^\top\Sigma^* P_{k*}^\top = 0,\end{aligned}$$

where in the second identity we use the fact that $\Sigma^*(I + \Theta^*\Sigma_{\mathcal{G}}^*) = \Sigma_{\mathcal{G}}^*$ and the fact that $\Sigma_{\mathcal{G}_2}^* P_{k*}^\top = \Sigma_{\mathcal{G}}^* P_{k*}^\top$ and in the third identity we use the fact that $(I - \Sigma^*\Theta^*)\Sigma_{\mathcal{G}}^* = \Sigma^*$.

For its conditional variance, we calculate it as follows.

$$\text{Var}(Z_{jk,i}|\mathcal{D}_2 = \mathbb{X}) = \underbrace{\mathbb{E}[(M_{j*}XX^\top(I + \Theta^*\Sigma_{\mathcal{G}}^*)P_{k*}^\top - M_{j*}(I - \Sigma^*\Theta^*)X_{\mathcal{G}_2}X_{\mathcal{G}_2}^\top P_{k*}^\top)^2]}_{I} - J,$$

where $J = (M_{j*}\Sigma_{\mathcal{G}}^* P_{k*}^\top - M_{j*}\Sigma^* P_{k*}^\top)^2$. For term $I$, we can decompose it into three terms.

$$\begin{aligned}I &= \underbrace{\mathbb{E}[M_{j*}XX^\top(I + \Theta^*\Sigma_{\mathcal{G}}^*)P_{k*}^\top P_{k*}(I + \Sigma_{\mathcal{G}}^*\Theta^*)XX^\top M_{j*}^\top]}_{I_1} \\ &+ \underbrace{\mathbb{E}[M_{j*}(I - \Sigma^*\Theta^*)X_{\mathcal{G}_2}X_{\mathcal{G}_2}^\top P_{k*}^\top P_{k*}(X_{\mathcal{G}_2}X_{\mathcal{G}_2}^\top)(I - \Theta^*\Sigma^*)M_{j*}^\top]}_{I_2} \\ &\underbrace{-2\mathbb{E}[M_{j*}XX^\top(I + \Theta^*\Sigma_{\mathcal{G}}^*)P_{k*}^\top M_{j*}(I - \Sigma^*\Theta^*)X_{\mathcal{G}_2}X_{\mathcal{G}_2}^\top P_{k*}^\top]}_{I_3}.\end{aligned} \quad (D.5)$$

In the following, we will extensively use Lemma D.4 to calculate $I_1, I_2$ and $I_3$. For $I_1$, we have

$$\begin{aligned}I_1 &= 2(M_{j*}\Sigma_{\mathcal{G}}^* P_{k*}^\top)[P_{k*}(I + \Sigma_{\mathcal{G}}^*\Theta^*)\Sigma^* M_{j*}^\top] + [P_{k*}(I + \Sigma_{\mathcal{G}}^*\Theta^*)\Sigma_{\mathcal{G}}^* P_{k*}^\top](M_{j*}\Sigma^* M_{j*}^\top) \\ &= 2(M_{j*}\Sigma_{\mathcal{G}}^* P_{k*}^\top)^2 + [P_{k*}(I + \Sigma_{\mathcal{G}}^*\Theta^*)\Sigma_{\mathcal{G}}^* P_{k*}^\top](M_{j*}\Sigma^* M_{j*}^\top),\end{aligned} \quad (D.6)$$

where we use $(I + \Sigma_{\mathcal{G}}^*\Theta^*)\Sigma^* = \Sigma_{\mathcal{G}}^*$ in the second equality. For $I_2$, we get

$$\begin{aligned}I_2 &= 2[M_{j*}(I - \Sigma^*\Theta^*)\Sigma_{\mathcal{G}}^* P_{k*}^\top]^2 + (P_{k*}\Sigma_{\mathcal{G}}^* P_{k*}^\top)[M_{j*}(I - \Sigma^*\Theta^*)\Sigma_{\mathcal{G}_2}^*(I - \Theta^*\Sigma^*)M_{j*}^\top] \\ &= 2(M_{j*}\Sigma^* P_{k*}^\top)^2 + (P_{k*}\Sigma_{\mathcal{G}}^* P_{k*}^\top)[M_{j*}(I - \Sigma^*\Theta^*)\Sigma_{\mathcal{G}_2}^*(I - \Theta^*\Sigma^*)M_{j*}^\top],\end{aligned} \quad (D.7)$$

where we use $(I - \Sigma^*\Theta^*)\Sigma_{\mathcal{G}}^* = \Sigma^*$ in the second equality. And for $I_3$, we have

$$\begin{aligned}I_3 = -2\{&(M_{j*}\Sigma_{\mathcal{G}}^* P_{k*}^\top)(M_{j*}\Sigma^* P_{k*}^\top) + [M_{j*}\Sigma^* I_2(I - \Theta^*\Sigma^*)M_{j*}^\top](P_{k*}\Sigma_{\mathcal{G}}^* P_{k*}^\top) \\ &+ [P_{k*}\Sigma_{\mathcal{G}}^* I_2(I - \Theta^*\Sigma^*)M_{j*}^\top](M_{j*}\Sigma^* P_{k*}^\top)\},\end{aligned} \quad (D.8)$$

where $I_2 = \text{diag}(0, I_{d_2})$. Combing (D.5), (D.6), (D.7) and (D.8), we have

$$\begin{aligned}\text{Var}(Z_{jk,i}|\mathcal{D}_2 = \mathbb{X}) &= I_1 + I_2 + I_3 \\ &= [P_{k*}(I + \Sigma_{\mathcal{G}}^*\Theta^*)\Sigma_{\mathcal{G}}^* P_{k*}^\top](M_{j*}\Sigma^* M_{j*}^\top) + (M_{j*}\Sigma_{\mathcal{G}}^* P_{k*}^\top)^2 \\ &- (M_{j*}\Sigma^* P_{k*}^\top)^2 - (P_{k*}\Sigma_{\mathcal{G}}^* P_{k*}^\top)[M_{j*}(I - \Sigma^*\Theta^*)\Sigma_{\mathcal{G}_2}^*(I - \Theta^*\Sigma^*)M_{j*}^\top].\end{aligned}$$

Thus we have $\text{Var}(Z_{jk,i}|\mathcal{D}_2 = \mathbb{X}) = \xi_{jk}^2$. Recall that conditional on $\mathcal{D} = \mathbb{X} \in \Gamma$, we have $|M_{j*}|_1 \leq \|M\|_\infty = \mathcal{O}(1)$ and $|P_{k*}|_1 \leq \|k\|_\infty = \mathcal{O}(1)$. Thus by Lemma D.5, we have for all $i \in [i]$, $\|Z_{jk,i}|\mathcal{D}_2 = \mathbb{X}\|_{\psi_1} = \mathcal{O}(1)$. By the definition of $\psi_1$-norm, we have

$$\rho = \mathbb{E}(|Z_{jk,i}|^3|\mathcal{D}_2 = \mathbb{X}) = \mathcal{O}(1).$$

Also, from the proof of Lemma D.3, we have $\xi_{jk}^2 \geq C'' > 0$ for some positive constant $C''$. Thus by Berry-Esseen Theorem, there exists a universal constant $C_1$ such that for every $\mathbb{X} \in \Gamma$ and for every $x \in \mathbb{R}$,

$$\left|\mathbb{P}\left(\frac{S_n}{s_n} \leq x | \mathcal{D}_2 = \mathbb{X}\right) - \Phi(x)\right| \leq \frac{C_1 \rho}{\xi_{jk}^3 \sqrt{n}} \leq \frac{C_2}{\sqrt{n}}, \tag{D.9}$$

where the second inequality comes from the fact that $\rho = \mathcal{O}(1)$ and $\xi_{jk} \geq C'' > 0$ and $C_2$ is also a universal constant. Define the event $\mathcal{E} = \{\mathcal{D}_2 \in \Gamma\}$. From Lemma E.5, we know that $\mathbb{P}(\mathcal{E}) \geq 1 - 2d^{-1}$. Using the law of total probability, we have for all $x \in \mathbb{R}$,

$$\left|\mathbb{P}\left(\frac{S_n}{s_n} \leq x\right) - \Phi(x)\right| = \left|\mathbb{P}\left(\frac{S_n}{s_n} \leq x | \mathcal{E}\right)\mathbb{P}(\mathcal{E}) + \mathbb{P}\left(\frac{S_n}{s_n} \leq x | \mathcal{E}^c\right)\mathbb{P}(\mathcal{E}^c) - \Phi(x)\right|$$

$$\leq \left|\left[\mathbb{P}\left(\frac{S_n}{s_n} \leq x | \mathcal{E}\right) - \Phi(x)\right]\mathbb{P}(\mathcal{E})\right| + \left|\left[\mathbb{P}\left(\frac{S_n}{s_n} \leq x | \mathcal{E}^c\right) - \Phi(x)\right]\mathbb{P}(\mathcal{E}^c)\right|$$

$$\leq \left|\mathbb{P}\left(\frac{S_n}{s_n} \leq x | \mathcal{E}\right) - \Phi(x)\right| + 2\mathbb{P}(\mathcal{E}^c),$$

where in the first inequality we use the triangle inequality and in the second one we use the fact that $|\mathbb{P}(\mathcal{E})| \leq 1$ and $|\mathbb{P}(S_n/s_n \leq x | \mathcal{E}^c) - \Phi(x)| \leq 2$. To continue, we have

$$\left|\mathbb{P}\left(\frac{S_n}{s_n} \leq x | \mathcal{E}\right) - \Phi(x)\right| = \left|\int \left[\mathbb{P}\left(\frac{S_n}{s_n} \leq x | \mathcal{E}, \mathcal{D}_2 = \mathbb{X}\right) - \Phi(x)\right] d\mathbb{P}\mathbb{X}\right| \leq \frac{C_2}{\sqrt{n}},$$

where the inequality comes from (D.9). Combining with $\mathbb{P}(\mathcal{E}^c) \leq 2d^{-1} \leq 2n^{-1}$, we have

$$\left|\mathbb{P}\left(\frac{S_n}{s_n} \leq x\right) - \Phi(x)\right| \leq \frac{C_2}{\sqrt{n}} + \frac{2}{n}.$$

This shows that $S_n/s_n \rightsquigarrow N(0,1)$. By (D.4) and Slutsky's theorem, we have $\sqrt{n} \cdot (\widehat{\theta}_{jk}^u - \theta_{jk}^*)/\xi_{jk} \rightsquigarrow N(0,1)$. This finishes the proof of Theorem 4.2. $\square$

Next we prove the variance estimator in (4.2) is consistent for (4.1).

*Proof of Lemma 4.4.* In light of Lemma D.3, it suffices to show that $\widehat{\xi}_{jk}^2 - \xi_{jk}^2 = o_\mathbb{P}(1)$. Recall from (4.1) and (4.2) that for $j \in [d_1]$, $k \in [d_1+1, d]$, we have

$$\xi_{jk}^2 = \underbrace{(M_{j*}\Sigma^* M_{j*}^\top)[P_{k*}(I + \Sigma_\mathcal{G}^*\Theta^*)\Sigma_\mathcal{G}^* P_{k*}^\top]}_{I_1} + \underbrace{(M_{j*}\Sigma_\mathcal{G}^* P_{k*}^\top)^2}_{I_2} - \underbrace{(M_{j*}\Sigma^* P_{k*}^\top)^2}_{I_3}$$

$$- \underbrace{[M_{j*}(I - \Sigma^*\Theta^*)\Sigma_{\mathcal{G}_2}^*(I - \Theta^*\Sigma^*)M_{j*}^\top](P_{k*}\Sigma_\mathcal{G}^* P_{k*}^\top)}_{I_4}.$$

And the estimator $\widehat{\xi}_{jk}^2$ is the empirical version of the above formula

$$\widehat{\xi}_{jk}^2 = \underbrace{(M_{j*}\widehat{\Sigma}M_{j*}^\top)[P_{k*}(I + \widehat{\Sigma}_{\mathcal{G}}\widehat{\Theta})\widehat{\Sigma}_{\mathcal{G}}P_{k*}^\top]}_{\widehat{I}_1} + \underbrace{(M_{j*}\widehat{\Sigma}_{\mathcal{G}}P_{k*}^\top)^2}_{\widehat{I}_2} - \underbrace{(M_{j*}\widehat{\Sigma}P_{k*}^\top)^2}_{\widehat{I}_3}$$
$$- \underbrace{[M_{j*}(I - \widehat{\Sigma}\widehat{\Theta})\widehat{\Sigma}_{\mathcal{G}_2}(I - \widehat{\Theta}\widehat{\Sigma})M_{j*}^\top](P_{k*}\widehat{\Sigma}_{\mathcal{G}}P_{k*}^\top)}_{\widehat{I}_4}.$$

We quantify the difference between $\xi_{jk}^2$ and $\widehat{\xi}_{jk}^2$ term by term. For the first term, we have

$$|I_1 - \widehat{I}_1| \leq \left|M_{j*}(\widehat{\Sigma} - \Sigma^*)M_{j*}^\top P_{k*}[(I + \widehat{\Sigma}_{\mathcal{G}}\widehat{\Theta})\widehat{\Sigma}_{\mathcal{G}} - (I + \Sigma_{\mathcal{G}}^*\Theta^*)\Sigma_{\mathcal{G}}^*]P_{k*}^\top\right|$$
$$+ \left|M_{j*}(\widehat{\Sigma} - \Sigma^*)M_{j*}^\top P_{k*}(I + \Sigma_{\mathcal{G}}^*\Theta^*)\Sigma_{\mathcal{G}}^*P_{k*}^\top\right|$$
$$+ \left|M_{j*}\Sigma^*M_{j*}^\top P_{k*}[(I + \widehat{\Sigma}_{\mathcal{G}}\widehat{\Theta})\widehat{\Sigma}_{\mathcal{G}} - (I + \Sigma_{\mathcal{G}}^*\Theta^*)\Sigma_{\mathcal{G}}^*]P_{k*}^\top\right|.$$

We will upper bound the right hand side using $|u^\top A v| \leq \|u\|_1 \|A\|_{\max} \|v\|_1$. By Lemma D.1, we have $\|M_{j*}\|_1 = \mathcal{O}_\mathbb{P}(1)$ and $\|P_{k*}\|_1 = \mathcal{O}_\mathbb{P}(1)$. Thus it remains to control $\|\widehat{\Sigma} - \Sigma^*\|_{\max}$, $\|(I + \widehat{\Sigma}_{\mathcal{G}}\widehat{\Theta})\widehat{\Sigma}_{\mathcal{G}} - (I + \Sigma_{\mathcal{G}}^*\Theta^*)\Sigma_{\mathcal{G}}^*\|_{\max}$, $\|(I + \Sigma_{\mathcal{G}}^*\Theta^*)\Sigma_{\mathcal{G}}^*\|_{\max}$ and $\|\Sigma^*\|_{\max}$. By Assumption (**A2**), we have $\|\Sigma^*\|_{\max} = \mathcal{O}(1)$ and by Lemma E.5, we have $\|\widehat{\Sigma} - \Sigma^*\|_{\max} = \mathcal{O}_\mathbb{P}(\sqrt{\log d/n})$. Using the fact that $\|ABC\|_{\max} \leq \|A\|_{\max}\|B\|_{1,1}\|C\|_{\max}$, we have

$$\|(I + \Sigma_{\mathcal{G}}^*\Theta^*)\Sigma_{\mathcal{G}}^*\|_{\max} \leq \|\Sigma_{\mathcal{G}}^*\|_{\max} + \|\Sigma_{\mathcal{G}}^*\Theta^*\Sigma_{\mathcal{G}}^*\|_{\max} \leq \|\Sigma_{\mathcal{G}}^*\|_{\max} + \|\Sigma_{\mathcal{G}}^*\|_{\max}\|\Theta^*\|_{1,1}\|\Sigma_{\mathcal{G}}^*\|_{\max}.$$

Thus $\|(I + \Sigma_{\mathcal{G}}^*\Theta^*)\Sigma_{\mathcal{G}}^*\|_{\max} = \mathcal{O}(1)$ by Assumptions (**A2**) and (**A3**). For its perturbation,

$$\|(I + \widehat{\Sigma}_{\mathcal{G}}\widehat{\Theta})\widehat{\Sigma}_{\mathcal{G}} - (I + \Sigma_{\mathcal{G}}^*\Theta^*)\Sigma_{\mathcal{G}}^*\|_{\max} \leq \|\widehat{\Sigma}_{\mathcal{G}} - \Sigma_{\mathcal{G}}^*\|_{\max} + \|\widehat{\Sigma}_{\mathcal{G}}\widehat{\Theta}\widehat{\Sigma}_{\mathcal{G}} - \Sigma_{\mathcal{G}}^*\Theta^*\Sigma_{\mathcal{G}}^*\|_{\max},$$

where we use the triangle inequality. $\|\widehat{\Sigma}_{\mathcal{G}} - \Sigma_{\mathcal{G}}^*\|_{\max} = \mathcal{O}_\mathbb{P}(\sqrt{\log d/n})$ by Lemma E.5. And for the remaining term $\|\widehat{\Sigma}_{\mathcal{G}}\widehat{\Theta}\widehat{\Sigma}_{\mathcal{G}} - \Sigma_{\mathcal{G}}^*\Theta^*\Sigma_{\mathcal{G}}^*\|_{\max}$, we have

$$\|\widehat{\Sigma}_{\mathcal{G}}\widehat{\Theta}\widehat{\Sigma}_{\mathcal{G}} - \Sigma_{\mathcal{G}}^*\Theta^*\Sigma_{\mathcal{G}}^*\|_{\max} \leq \|(\widehat{\Sigma}_{\mathcal{G}} - \Sigma_{\mathcal{G}}^*)(\widehat{\Theta} - \Theta^*)(\widehat{\Sigma}_{\mathcal{G}} - \Sigma_{\mathcal{G}}^*)\|_{\max}$$
$$+ \|(\widehat{\Sigma}_{\mathcal{G}} - \Sigma_{\mathcal{G}}^*)(\widehat{\Theta} - \Theta^*)\Sigma_{\mathcal{G}}^*\|_{\max} + \|(\widehat{\Sigma}_{\mathcal{G}} - \Sigma_{\mathcal{G}}^*)\Theta^*(\widehat{\Sigma}_{\mathcal{G}} - \Sigma_{\mathcal{G}}^*)\|_{\max} + \|(\widehat{\Sigma}_{\mathcal{G}} - \Sigma_{\mathcal{G}}^*)\Theta^*\Sigma_{\mathcal{G}}^*\|_{\max}$$
$$+ \|\Sigma_{\mathcal{G}}^*(\widehat{\Theta} - \Theta^*)(\widehat{\Sigma}_{\mathcal{G}} - \Sigma_{\mathcal{G}}^*)\|_{\max} + \|\Sigma_{\mathcal{G}}^*(\widehat{\Theta} - \Theta^*)\Sigma_{\mathcal{G}}^*\|_{\max} + \|\Sigma_{\mathcal{G}}^*\Theta^*(\widehat{\Sigma}_{\mathcal{G}} - \Sigma_{\mathcal{G}}^*)\|_{\max}.$$

By extensively using $\|ABC\|_{\max} \leq \|A\|_{\max}\|B\|_{1,1}\|C\|_{\max}$, Lemma E.5 and Theorem 4.1, we have $\|\widehat{\Sigma}_{\mathcal{G}}\widehat{\Theta}\widehat{\Sigma}_{\mathcal{G}} - \Sigma_{\mathcal{G}}^*\Theta^*\Sigma_{\mathcal{G}}^*\|_{\max} = \mathcal{O}_\mathbb{P}(s^*\sqrt{\log d/n})$. In all, we have $|I_1 - \widehat{I}_1| = \mathcal{O}_\mathbb{P}(s^*\sqrt{\log d/n})$.

The second term can be bounded as follows. By triangle inequality, we have

$$|I_2 - \widehat{I}_2| \leq |[M_{j*}(\widehat{\Sigma}_{\mathcal{G}} - \Sigma_{\mathcal{G}}^*)P_{k*}^\top]^2| + 2|[M_{j*}(\widehat{\Sigma}_{\mathcal{G}} - \Sigma_{\mathcal{G}}^*)P_{k*}^\top](M_{j*}\Sigma_{\mathcal{G}}^*P_{k*}^\top)|.$$

Again, using $|u^\top A v| \leq \|u\|_1 \|A\|_{\max} \|v\|_1$, we have $|I_2 - \widehat{I}_2| = \mathcal{O}_\mathbb{P}(\sqrt{\log d/n})$, where we use $\|M_{j*}\|_1 = \mathcal{O}_\mathbb{P}(1)$, $\|P_{k*}\|_1 = \mathcal{O}_\mathbb{P}(1)$, $\|\widehat{\Sigma}_{\mathcal{G}} - \Sigma_{\mathcal{G}}^*\|_{\max} = \mathcal{O}_\mathbb{P}(\sqrt{\log d/n})$ and $\|\Sigma_{\mathcal{G}}^*\|_{\max} = \mathcal{O}(1)$.

Similar proofs can show that $|I_3 - \widehat{I}_3| = \mathcal{O}_\mathbb{P}(\sqrt{\log d/n})$ and $|I_4 - \widehat{I}_4| = \mathcal{O}_\mathbb{P}(s^*\sqrt{\log d/n})$. In all, we show that $|\widehat{\xi}_{jk}^2 - \xi_{jk}^2| = \mathcal{O}_\mathbb{P}(s^*\sqrt{\log d/n}) = o_\mathbb{P}(1)$. $\square$

# E  Auxiliary Lemmas for Estimation Consistency

In this section, we give detailed proofs of the technical lemmas provided in Section C. We first review basic definitions of sub-Gaussian and sub-exponential random variables.

**Definition E.1.** (Sub-Gaussian random variable and sub-Gaussian norm) A random variable $X$ is sub-Gaussian if there exists a constant $K_1$ such that $\mathbb{P}(|X| \geq t) \leq \exp(1 - t^2/K_1^2)$ for all $t \geq 0$. The sub-Gaussian norm of $X$ is defined as $\|X\|_{\psi_2} = \sup_{p \geq 1} p^{-1/2} (\mathbb{E}|X|^p)^{1/p}$.

**Definition E.2.** (Sub-exponential random variable and sub-exponential norm) A random variable $X$ is sub-exponential if there exists a constant $K_2$ such that $\mathbb{P}(|X| \geq t) \leq \exp(1 - t/K_2)$ for all $t \geq 0$. The sub-exponential norm of $X$ is defined as $\|X\|_{\psi_1} = \sup_{p \geq 1} p^{-1} (\mathbb{E}|X|^p)^{1/p}$.

The next two lemmas present the basic properties of the sub-Gaussian and sub-exponential random variables.

**Lemma E.3.** (Product of Sub-Gaussians) Assume that $X$ and $Y$ are sub-Gaussian random variables, then we have $\|XY\|_{\psi_1} \leq 2\|X\|_{\psi_2}\|Y\|_{\psi_2}$.

*Proof.* The proof of this can be found in Lemma 6.2 in Javanmard and Montanari (2014). □

**Lemma E.4.** (Bernstein's Inequality) Let $X_1, \ldots, X_n$ be independent centered sub-exponential random variables, and $B = \max_i \|X_i\|_{\psi_1}$. Then for any $t > 0$, we have

$$\mathbb{P}\Big(\Big|\frac{1}{n}\sum_{i=1}^n X_i\Big| \geq t\Big) \leq 2\exp\Big[-c\min\Big(\frac{t^2}{B^2}, \frac{t}{B}\Big)n\Big],$$

where $c > 0$ is an absolute constant.

*Proof.* See Proposition 5.16 in Vershynin (2010) for a detailed proof of this. □

Lemma E.5 gives an application of Lemma E.4 to the estimation of covariance matrices.

**Lemma E.5.** (Sample Covariance Matrix) Recall that $\widehat{\Sigma} = (1/n) \sum_{i=1}^n \mathbb{X}^\top \mathbb{X}$, we have

$$\mathbb{P}\big(\|\widehat{\Sigma} - \Sigma^*\|_{\max} \geq t\big) \leq 2d^2 \exp\Big[-c\min\Big(\frac{t^2}{K^2}, \frac{t}{K}\Big)n\Big],$$

where $c$ is an absolute constant and $K$ is the upper bound specified in Assumption (**A2**).

*Proof.* By union bound, we have $\mathbb{P}\big(\|\widehat{\Sigma} - \Sigma^*\|_{\max} \geq t\big) \leq d^2 \max_{j,k} \mathbb{P}\big(|\widehat{\Sigma}_{jk} - \sigma_{jk}^*| \geq t\big)$, where $\sigma_{jk}^*$ is the $(j,k)$-th entry in $\Sigma^*$. For any $j \in [d]$ and $k \in [d]$, we have

$$\mathbb{P}\big(|\widehat{\Sigma}_{jk} - \sigma_{jk}^*| \geq t\big) = \mathbb{P}\Big(\Big|\frac{1}{n}\sum_{i=1}^n X_{ij}X_{ik} - \sigma_{jk}^*\Big| \geq t\Big),$$

where $X_{ij}$ denote the $j$-th variable in the $i$-th sample. By Assumption (**A2**), we have $\|\Sigma^*\|_{\max} \leq K$, which means $\max_{1 \leq j \leq d} \text{Var}(X_{ij}) \leq K$. Thus $\{X_{ij}\}_{j=1}^d$ are sub-Gaussian random variables with $\max_j \|X_{ij}\|_{\psi_2} \leq C\sqrt{K}$, where $C$ is an absolute constant. By Lemma E.3, we have $\|X_{ij}X_{ik}\|_{\psi_1} \leq 2C^2K$ and $\|X_{ij}X_{ik} - \sigma_{jk}^*\|_{\psi_1} \leq 4C^2K$ by the centering property of sub-exponential norm (Vershynin, 2010). By the Bernstein's inequality in Lemma E.4,

$$\mathbb{P}\Big(\Big|\frac{1}{n}\sum_{i=1}^n X_{ij}X_{ik} - \sigma_{jk}^*\Big| \geq t\Big) \leq 2\exp\Big[-c\min\Big(\frac{t^2}{K^2}, \frac{t}{K}\Big)n\Big], \tag{E.1}$$

where $c$ is a constant. Combing with the union bound, we prove the tail bound for $\widehat{\Sigma}$. $\square$

Lemma E.6 provides perturbation bound for the inverse of a matrix.

**Lemma E.6.** (Matrix Inverse Perturbation) Let $A \in \mathbb{R}^{d \times d}$ be the target matrix and $E \in \mathbb{R}^{d \times d}$ be the perturbation matrix. If $\|A^{-1}\|_1 \|E\|_1 \leq 1/2$, then we have

$$\|(A+E)^{-1} - A^{-1}\|_1 \leq 2\|A^{-1}\|_1^2 \cdot \|E\|_1.$$

*Proof.* See Fact 2.25 in Ipsen (2009) for a detailed proof of this. $\square$

### E.1  $\ell_{\max}$-norm Bound

*Proof of Lemma C.1.* Recall that the empirical loss function is defined as

$$\mathcal{L}_n(\Theta) = \text{Tr}(\Theta\widehat{\Sigma}) - \log|\widehat{\Sigma}_{\mathcal{G}}\Theta\widehat{\Sigma}_{\mathcal{G}} + \widehat{\Sigma}_{\mathcal{G}}|.$$

Thus the gradient can be calculated as follows

$$\nabla \mathcal{L}_n(\Theta) = \widehat{\Sigma} - \widehat{\Sigma}_{\mathcal{G}}(\Theta\widehat{\Sigma}_{\mathcal{G}} + I)^{-1}. \tag{E.2}$$

Substituting $\Theta^*$ into (E.2), we get $\nabla \mathcal{L}_n(\Theta^*) = \widehat{\Sigma} - \widehat{\Sigma}_{\mathcal{G}}(\Theta^*\widehat{\Sigma}_{\mathcal{G}} + I)^{-1}$. Thus we have

$$\|\nabla \mathcal{L}_n(\Theta^*)\|_{\max} = \|[\widehat{\Sigma} - \widehat{\Sigma}_{\mathcal{G}}(\Theta^*\widehat{\Sigma}_{\mathcal{G}} + I)^{-1}](\Theta^*\widehat{\Sigma}_{\mathcal{G}} + I)(\Theta^*\widehat{\Sigma}_{\mathcal{G}} + I)^{-1}\|_{\max}$$
$$= \|(\widehat{\Sigma} + \widehat{\Sigma}\Theta^*\widehat{\Sigma}_{\mathcal{G}} - \widehat{\Sigma}_{\mathcal{G}})(\Theta^*\widehat{\Sigma}_{\mathcal{G}} + I)^{-1}\|_{\max},$$

where in the first equality we use the identity $(\Theta^*\widehat{\Sigma}_{\mathcal{G}} + I)(\Theta^*\widehat{\Sigma}_{\mathcal{G}} + I)^{-1} = I$. Further using the inequality $\|AB\|_{\max} \leq \|A\|_{\max}\|B\|_1$, we get

$$\|\nabla \mathcal{L}_n(\Theta^*)\|_{\max} \leq \underbrace{\|\widehat{\Sigma} + \widehat{\Sigma}\Theta^*\widehat{\Sigma}_{\mathcal{G}} - \widehat{\Sigma}_{\mathcal{G}}\|_{\max}}_{I_1} \cdot \underbrace{\|(\Theta^*\widehat{\Sigma}_{\mathcal{G}} + I)^{-1}\|_1}_{I_2}. \tag{E.3}$$

For the first term, since $\Sigma^* + \Sigma^*\Theta^*\Sigma_{\mathcal{G}}^* - \Sigma_{\mathcal{G}}^* = 0$, we have

$$I_1 = \|\widehat{\Sigma} + \widehat{\Sigma}\Theta^*\widehat{\Sigma}_{\mathcal{G}} - \widehat{\Sigma}_{\mathcal{G}} - (\Sigma^* + \Sigma^*\Theta^*\Sigma_{\mathcal{G}}^* - \Sigma_{\mathcal{G}}^*)\|_{\max}$$
$$\leq \|\widehat{\Sigma} - \Sigma^*\|_{\max} + \|\widehat{\Sigma}_{\mathcal{G}} - \Sigma_{\mathcal{G}}^*\|_{\max} + \|(\widehat{\Sigma} - \Sigma^*)\Theta^*(\widehat{\Sigma}_{\mathcal{G}} - \Sigma_{\mathcal{G}}^*)\|_{\max}$$
$$\quad + \|(\widehat{\Sigma} - \Sigma^*)\Theta^*\Sigma_{\mathcal{G}}^*\|_{\max} + \|\Sigma^*\Theta^*(\widehat{\Sigma}_{\mathcal{G}} - \Sigma_{\mathcal{G}}^*)\|_{\max}$$
$$\leq \|\widehat{\Sigma} - \Sigma^*\|_{\max} + \|\widehat{\Sigma}_{\mathcal{G}} - \Sigma_{\mathcal{G}}^*\|_{\max} + \|\widehat{\Sigma} - \Sigma^*\|_{\max}\|\Theta^*\|_{1,1}\|\widehat{\Sigma}_{\mathcal{G}} - \Sigma_{\mathcal{G}}^*\|_{\max}$$
$$\quad + \|\widehat{\Sigma} - \Sigma^*\|_{\max}\|\Theta^*\|_{1,1}\|\Sigma_{\mathcal{G}}^*\|_{\max} + \|\Sigma^*\|_{\max}\|\Theta^*\|_{1,1}\|\widehat{\Sigma}_{\mathcal{G}} - \Sigma_{\mathcal{G}}^*\|_{\max}, \tag{E.4}$$

where in the first inequality we use the triangle inequality and in the second inequality we use the fact that $\|ABC\|_{\max} \leq \|A\|_{\max}\|B\|_{1,1}\|C\|_{\max}$.

By Lemma E.5, there exists a constant $C_1$ such that for the event $\mathcal{E} = \{\|\widehat{\Sigma} - \Sigma^*\|_{\max} \leq C_1\sqrt{\log d/n}\}$, we have $\mathbb{P}(\mathcal{E}) \geq 1 - 2d^{-1}$. In the following, we are always conditioning on $\mathcal{E}$. Under the event $\mathcal{E}$, we also have $\|\widehat{\Sigma}_{\mathcal{G}} - \Sigma^*_{\mathcal{G}}\|_{\max} \leq C_1\sqrt{\log d/n}$ by the definitions of $\widehat{\Sigma}_{\mathcal{G}}$ and $\Sigma^*_{\mathcal{G}}$. Considering (E.4), we have

$$I_1 \leq 2C_1\sqrt{\frac{\log d}{n}} + C_1^2 R\frac{\log d}{n} + 2KRC_1\sqrt{\frac{\log d}{n}} \leq C_2\sqrt{\frac{\log d}{n}}, \tag{E.5}$$

where we use Assumptions (**A2**) and (**A3**) and the fact that $\sqrt{\log d/n} = o(1)$.

For the second term, by triangle inequality we have

$$I_2 \leq \underbrace{\|(\Theta^*\widehat{\Sigma}_{\mathcal{G}} + I)^{-1} - (\Theta^*\Sigma^*_{\mathcal{G}} + I)^{-1}\|_1}_{I_{2,1}} + \underbrace{\|(\Theta^*\Sigma^*_{\mathcal{G}} + I)^{-1}\|_1}_{I_{2,2}}. \tag{E.6}$$

$I_{2,2}$ is a population quantity which we can bound easily as follows.

$$I_{2,2} = \|I - \Theta^*\Sigma^*\|_1 \leq \|I\|_1 + \|\Theta^*\Sigma^*\|_1 \leq \|I\|_1 + \|\Theta^*\|_{1,1}\|\Sigma^*\|_{\max} \leq 1 + KR \leq C_3, \tag{E.7}$$

where in the first inequality we use the triangle inequality, in the second we use the fact that $\|AB\|_1 \leq \|A\|_{1,1}\|B\|_{\max}$ and in the third we use Assumptions (**A2**) and (**A3**) again. $I_{2,1}$ is the perturbation of the matrix inverse. Denote $A = \Theta^*\Sigma^*_{\mathcal{G}} + I$ and $E = \Theta^*(\widehat{\Sigma}_{\mathcal{G}} - \Sigma^*_{\mathcal{G}})$. We first check the condition in Lemma E.6. We have

$$\|A^{-1}\|_1\|E\|_1 \leq \|(\Theta^*\Sigma^*_{\mathcal{G}} + I)^{-1}\|_1\|\Theta^*\|_{1,1}\|(\widehat{\Sigma}_{\mathcal{G}} - \Sigma^*_{\mathcal{G}})\|_{\max} \leq C_1C_3M\sqrt{\frac{\log d}{n}} \leq C_4\sqrt{\frac{\log d}{n}},$$

where we use (E.7), the inequality $\|AB\|_1 \leq \|A\|_{1,1}\|B\|_{\max}$ and the fact that $\|\widehat{\Sigma}_{\mathcal{G}} - \Sigma^*_{\mathcal{G}}\|_{\max} \leq C_1\sqrt{\log d/n}$. Since $\sqrt{\log d/n} = o(1)$, we have $\|A^{-1}\|_1\|E\|_1 \leq 1/2$, thus by Lemma E.6,

$$I_{2,1} \leq 2\|(\Theta^*\Sigma^*_{\mathcal{G}} + I)^{-1}\|_1^2 \cdot \|\Theta^*(\widehat{\Sigma}_{\mathcal{G}} - \Sigma^*_{\mathcal{G}})\|_1 \leq 2C_1C_3^2R\sqrt{\frac{\log d}{n}} \leq C_5\sqrt{\frac{\log d}{n}}. \tag{E.8}$$

Combining (E.6), (E.7) and (E.8), we have

$$I_2 \leq C_5\sqrt{\frac{\log d}{n}} + C_3 \leq 2C_3, \tag{E.9}$$

where we use the fact that $\sqrt{\log d/n} = o(1)$. Using (E.3), (E.5) and (E.9), we have with probability at least $1 - 2d^{-1}$,

$$\|\nabla \mathcal{L}_n(\Theta^*)\|_{\max} \leq 2C_2C_3\sqrt{\frac{\log d}{n}} \leq C\sqrt{\frac{\log d}{n}}.$$

Hence we prove the lemma. $\square$

## E.2 Restricted Strong Convexity

*Proof of Lemma C.2.* By taking derivatives, we have

$$\nabla^2 \mathcal{L}_n(\Theta) = [(\widehat{\Sigma}_\mathcal{G}\Theta + I)^{-1}\widehat{\Sigma}_\mathcal{G}] \otimes [(\widehat{\Sigma}_\mathcal{G}\Theta + I)^{-1}\widehat{\Sigma}_\mathcal{G}] \quad \text{and}$$
$$\nabla^2 \mathcal{L}(\Theta) = [(\Sigma_\mathcal{G}^*\Theta + I)^{-1}\Sigma_\mathcal{G}^*] \otimes [(\Sigma_\mathcal{G}^*\Theta + I)^{-1}\Sigma_\mathcal{G}^*].$$

Since $(\Sigma_\mathcal{G}^*\Theta^* + I)^{-1}\Sigma_\mathcal{G}^* = \Sigma^*$, we have $\nabla^2 \mathcal{L}(\Theta^*) = \Sigma^* \otimes \Sigma^*$. For all $\Delta \in \mathbb{C}^{(t^*)}$ and $\mu \in [0, 1]$, denote $\widetilde{\Delta} = \mu\Delta$, $\widetilde{\Sigma} = (\widehat{\Sigma}_\mathcal{G}\Theta^* + I)^{-1}\widehat{\Sigma}_\mathcal{G}$ and $\widetilde{\Sigma}' = [\widehat{\Sigma}_\mathcal{G}(\Theta^* + \widetilde{\Delta}) + I]^{-1}\widehat{\Sigma}_\mathcal{G}$, we have

$$\|\nabla^2 \mathcal{L}_n(\Theta^* + \widetilde{\Delta}) - \nabla^2 \mathcal{L}(\Theta^*)\|_{\max} = \|\widetilde{\Sigma}' \otimes \widetilde{\Sigma}' - \Sigma^* \otimes \Sigma^*\|_{\max}$$
$$\leq \underbrace{\|\widetilde{\Sigma}' \otimes \widetilde{\Sigma}' - \widetilde{\Sigma} \otimes \widetilde{\Sigma}\|_{\max}}_{I_1} + \underbrace{\|\widetilde{\Sigma} \otimes \widetilde{\Sigma} - \Sigma^* \otimes \Sigma^*\|_{\max}}_{I_2}, \quad (E.10)$$

where we use the triangle inequality. For the second term, we have

$$I_2 = \|(\widetilde{\Sigma} - \Sigma^* + \Sigma^*) \otimes (\widetilde{\Sigma} - \Sigma^* + \Sigma^*) - \Sigma^* \otimes \Sigma^*\|_{\max}$$
$$= \|(\widetilde{\Sigma} - \Sigma^*) \otimes (\widetilde{\Sigma} - \Sigma^*) + (\widetilde{\Sigma} - \Sigma^*) \otimes \Sigma^* + \Sigma^* \otimes (\widetilde{\Sigma} - \Sigma^*)\|_{\max},$$

where we use the fact that Kronecker product is bilinear. By triangle inequality and the fact that $\|A \otimes B\|_{\max} \leq \|A\|_{\max}\|B\|_{\max}$, we have

$$I_2 \leq \|\widetilde{\Sigma} - \Sigma^*\|_{\max}^2 + 2\|\Sigma^*\|_{\max} \cdot \|\widetilde{\Sigma} - \Sigma^*\|_{\max}. \quad (E.11)$$

Thus we need to control $\|\widetilde{\Sigma} - \Sigma^*\|_{\max}$. For this, we have

$$\|\widetilde{\Sigma} - \Sigma^*\|_{\max} = \|(\widehat{\Sigma}_\mathcal{G}\Theta^* + I)^{-1}\widehat{\Sigma}_\mathcal{G} - (\Sigma_\mathcal{G}^*\Theta^* + I)^{-1}\Sigma_\mathcal{G}^*\|_{\max}$$
$$\leq \|[(\widehat{\Sigma}_\mathcal{G}\Theta^* + I)^{-1} - (\Sigma_\mathcal{G}^*\Theta^* + I)^{-1}]\widehat{\Sigma}_\mathcal{G}\|_{\max} + \|(\Sigma_\mathcal{G}^*\Theta^* + I)^{-1}(\widehat{\Sigma}_\mathcal{G} - \Sigma_\mathcal{G}^*)\|_{\max},$$

where we use the triangle inequality. By $\|AB\|_{\max} \leq \|A\|_\infty \|B\|_{\max} = \|A^\top\|_1 \|B\|_{\max}$, we get

$$\|\widetilde{\Sigma} - \Sigma^*\|_{\max} \leq \|(\Theta^*\widehat{\Sigma}_\mathcal{G} + I)^{-1} - (\Theta^*\Sigma_\mathcal{G}^* + I)^{-1}\|_1 \|\widehat{\Sigma}_\mathcal{G}\|_{\max} + \|(\Theta^*\Sigma_\mathcal{G}^* + I)^{-1}\|_1 \|(\widehat{\Sigma}_\mathcal{G} - \Sigma_\mathcal{G}^*)\|_{\max}.$$

These four terms appear in the proof of Lemma C.1. Define the same event $\mathcal{E} = \{\|\widehat{\Sigma} - \Sigma^*\|_{\max} \leq C_1\sqrt{\log d/n}\}$ and by Lemma E.5, we have $\mathbb{P}(\mathcal{E}) \geq 1 - 2d^{-1}$. The following arguments are all conditioning on the event $\mathcal{E}$.

From the proof of Lemma C.1, specifically from (E.7) and (E.8), we have $\|(\Theta^*\Sigma_\mathcal{G}^* + I)^{-1}\|_1 \leq C_3$ and $\|(\Theta^*\widehat{\Sigma}_\mathcal{G} + I)^{-1} - (\Theta^*\Sigma_\mathcal{G}^* + I)^{-1}\|_1 \leq C_5\sqrt{\log d/n}$. Further, we have

$$\|\widehat{\Sigma}_\mathcal{G}\|_{\max} \leq \|\widehat{\Sigma}_\mathcal{G} - \Sigma_\mathcal{G}^*\|_{\max} + \|\Sigma_\mathcal{G}^*\|_{\max} \leq \|\widehat{\Sigma} - \Sigma^*\|_{\max} + \|\Sigma^*\|_{\max} \leq C_1\sqrt{\frac{\log d}{n}} + K \leq 2K,$$

where we use the definitions of $\widehat{\Sigma}_\mathcal{G}$ and $\Sigma_\mathcal{G}^*$ and the condition that $\sqrt{\log d/n} = o(1)$. Combining these together, we have

$$\|\widetilde{\Sigma} - \Sigma^*\|_{\max} \leq 2C_5 K\sqrt{\frac{\log d}{n}} + C_3 C_1 \sqrt{\frac{\log d}{n}} \leq C_6 \sqrt{\frac{\log d}{n}}. \quad (E.12)$$

Putting (E.12) into (E.11) and using $\sqrt{\log d/n} = o(1)$, we get

$$I_2 \leq C_6^2 \cdot \frac{\log d}{n} + 2K\sqrt{\frac{\log d}{n}} \leq C_7 \sqrt{\frac{\log d}{n}}. \tag{E.13}$$

For the first term in (E.10), we follow the same strategy as above. We have

$$\begin{aligned} I_1 &= \|(\widetilde{\Sigma}' - \widetilde{\Sigma} + \widetilde{\Sigma}) \otimes (\widetilde{\Sigma}' - \widetilde{\Sigma} + \widetilde{\Sigma}) - \widetilde{\Sigma} \otimes \widetilde{\Sigma}\|_{\max} \\ &= \|(\widetilde{\Sigma}' - \widetilde{\Sigma}) \otimes (\widetilde{\Sigma}' - \widetilde{\Sigma}) + (\widetilde{\Sigma}' - \widetilde{\Sigma}) \otimes \widetilde{\Sigma} + \widetilde{\Sigma} \otimes (\widetilde{\Sigma}' - \widetilde{\Sigma})\|_{\max} \\ &\leq \|\widetilde{\Sigma}' - \widetilde{\Sigma}\|_{\max}^2 + 2\|\widetilde{\Sigma}\|_{\max}\|\widetilde{\Sigma}' - \widetilde{\Sigma}\|_{\max}. \end{aligned} \tag{E.14}$$

Thus we need to control $\|\widetilde{\Sigma}\|_{\max}$ and $\|\widetilde{\Sigma}' - \widetilde{\Sigma}\|_{\max}$. By (E.12) and triangle inequality, we have

$$\|\widetilde{\Sigma}\|_{\max} \leq \|\widetilde{\Sigma} - \Sigma^*\|_{\max} + \|\Sigma^*\|_{\max} \leq C_6\sqrt{\frac{\log d}{n}} + K \leq 2K.$$

Use the fact that $\|AB\|_{\max} \leq \|A\|_\infty \|B\|_{\max} = \|A^\top\|_1 \|B\|_{\max}$, we also have

$$\begin{aligned} \|\widetilde{\Sigma}' - \widetilde{\Sigma}\|_{\max} &= \|(\widehat{\Sigma}_{\mathcal{G}}(\Theta^* + \widetilde{\Delta}) + I)^{-1}\widehat{\Sigma}_{\mathcal{G}} - (\widehat{\Sigma}_{\mathcal{G}}\Theta^* + I)^{-1}\widehat{\Sigma}_{\mathcal{G}}\|_{\max} \\ &\leq \|[(\Theta^* + \widetilde{\Delta})\widehat{\Sigma}_{\mathcal{G}} + I]^{-1} - (\Theta^*\widehat{\Sigma}_{\mathcal{G}} + I)^{-1}\|_1 \|\widehat{\Sigma}_{\mathcal{G}}\|_{\max}. \end{aligned} \tag{E.15}$$

We know that $\|\widehat{\Sigma}_{\mathcal{G}}\|_{\max} \leq 2K$. For the other term, we need to use Lemma E.6. let $A = \Theta^* \widehat{\Sigma}_{\mathcal{G}} + I$ and $E = \widetilde{\Delta}\widehat{\Sigma}_{\mathcal{G}}$. We first need to check the condition of Lemma E.6. Since

$$\|A^{-1}\|_1 \|E\|_1 = \|(\Theta^*\widehat{\Sigma}_{\mathcal{G}} + I)^{-1}\|_1 \|\widetilde{\Delta}\widehat{\Sigma}_{\mathcal{G}}\|_1 \leq 4C_3 K \|\widetilde{\Delta}\|_{1,1},$$

where we use the same argument as in the proof of Lemma C.1. Since $\Delta \in \mathbb{C}$, we have $\|\Delta_{S^c}\|_{1,1} \leq 3\|\Delta_S\|_{1,1}$. And this leads to

$$\|\Delta\|_{1,1} \leq 4\|\Delta_S\|_{1,1} \leq 4\sqrt{s^*}\|\Delta_S\|_F \leq 4\sqrt{s^*}\|\Delta\|_F, \tag{E.16}$$

where in the second inequality we use the Holder's inequality and in the third we use the fact that $\|\Delta_S\|_F \leq \|\Delta\|_F$. Further because $\|\Delta\|_F = t^*$ and $\mu \in [0,1]$, we have

$$\|\widetilde{\Delta}\|_{1,1} = \|\mu\Delta\|_{1,1} \leq \|\Delta\|_{1,1} = \mathcal{O}\left(s^*\sqrt{\frac{\log d}{n}}\right).$$

Therefore we have $\|A^{-1}\|_1 \|E\|_1 = o(1) < 1/2$, and

$$\|[(\Theta^* + \widetilde{\Delta})\widehat{\Sigma}_{\mathcal{G}} + I]^{-1} - (\Theta^*\widehat{\Sigma}_{\mathcal{G}} + I)^{-1}\|_1 \leq 2\|I - \Theta^*\widehat{\Sigma}_{\mathcal{G}}\|_1^2 \cdot \|\widetilde{\Delta}\widehat{\Sigma}_{\mathcal{G}}\|_1 = \mathcal{O}\left(s^*\sqrt{\frac{\log d}{n}}\right).$$

As a result, by (E.15), we have $\|\widetilde{\Sigma}' - \widetilde{\Sigma}\|_{\max} = \mathcal{O}(s^*\sqrt{\log d/n})$. Further yy (E.14), we have $I_1 = \mathcal{O}(s^*\sqrt{\log d/n})$. Combining it with (E.13), we have

$$\|\nabla^2 \mathcal{L}_n(\Theta^* + \widetilde{\Delta}) - \nabla^2 \mathcal{L}(\Theta^*)\|_{\max} = \mathcal{O}\left(s^*\sqrt{\frac{\log d}{n}}\right).$$

For any $\Delta$, we have $|\text{vec}(\Delta)^\top \big(\nabla^2 \mathcal{L}_n(\Theta^* + \widetilde{\Delta}) - \nabla^2 \mathcal{L}(\Theta^*)\big)\text{vec}(\Delta)| \leq \|\Delta\|_{1,1}^2 \|\nabla^2 \mathcal{L}_n(\Theta^* + \widetilde{\Delta}) - \nabla^2 \mathcal{L}_n(\Theta^*)\|_{\max}$, where we use the fact that $|x^\top A x| \leq \|x\|_1^2 \|A\|_{\max}$ and $\|\text{vec}(\Delta)\|_1 = \|\Delta\|_{1,1}$. Moreover, by (E.16) we have $\|\Delta\|_{1,1} \leq 4\sqrt{s^*}\|\Delta\|_F$. Combining all, we have

$$|\text{vec}(\Delta)^\top [\nabla^2 \mathcal{L}_n(\Theta^* + \widetilde{\Delta}) - \nabla^2 \mathcal{L}(\Theta^*)] \text{vec}(\Delta)| = \mathcal{O}\Big(s^{*2} \sqrt{\frac{\log d}{n}} \|\Delta\|_F^2\Big).$$

By triangle inequality, we also have

$$|\text{vec}(\Delta)^\top \nabla^2 \mathcal{L}(\Theta^*) \text{vec}(\Delta) - \text{vec}(\Delta)^\top \nabla^2 \mathcal{L}_n(\Theta^* + \widetilde{\Delta}) \text{vec}(\Delta)|$$
$$\geq |\text{vec}(\Delta)^\top \nabla^2 \mathcal{L}(\Theta^*) \text{vec}(\Delta)| - |\text{vec}(\Delta)^\top \nabla^2 \mathcal{L}_n(\Theta^* + \widetilde{\Delta}) \text{vec}(\Delta)|$$
$$= |\text{vec}(\Delta)^\top (\Sigma^* \otimes \Sigma^*) \text{vec}(\Delta)| - |\text{vec}(\Delta)^\top \nabla^2 \mathcal{L}_n(\Theta^* + \widetilde{\Delta}) \text{vec}(\Delta)|$$
$$\geq \rho_{\min}^2 \|\Delta\|_F^2 - |\text{vec}(\Delta)^\top \nabla^2 \mathcal{L}_n(\Theta^* + \widetilde{\Delta}) \text{vec}(\Delta)|,$$

where the first equality follows from the fact that $\nabla^2 \mathcal{L}(\Theta^*) = \Sigma^* \otimes \Sigma^*$ and in the second inequality, we use Assumption (**A4**) and the fact that the eigenvalues of Kronecker products of symmetric matrices are the products of the eigenvalues of the two matrices. Hence

$$|\text{vec}(\Delta)^\top \nabla^2 \mathcal{L}_n(\Theta^* + \widetilde{\Delta}) \text{vec}(\Delta)| \geq \Big(\rho_{\min}^2 - C' s^{*2} \sqrt{\frac{\log d}{n}}\Big) \|\Delta\|_F^2.$$

Thus we finish the proof. □

## F Auxillary Lemmas for the De-biased Estimator

In this section, we give the proof of the technical lemmas in Section D. For notational simplicity, denote $\Omega_{\mathcal{G}}^* = (\Sigma_{\mathcal{G}}^*)^{-1}$. First, we will state several useful lemmas.

The following two lemmas quantifies the size of $\Omega^* \widehat{\Sigma}' - I$ and $\Omega_{\mathcal{G}}^* \widehat{\Sigma}_{\mathcal{G}}' - I$.

**Lemma F.1.** (Feasibility of $\Omega^*$) Recall that $\widehat{\Sigma}'$ is the sample covariance matrix of the second sample $\mathcal{D}_2$, we have $\|\Omega^* \widehat{\Sigma}' - I\|_{\max} = \mathcal{O}_\mathbb{P}(\sqrt{\log d/n})$.

*Proof.* Since $\Omega^* \widehat{\Sigma}' - I = \Omega^*(\widehat{\Sigma}' - \Sigma^*)$, we have $\|\Omega^* \widehat{\Sigma}' - I\|_{\max} \leq \|\Omega^*\|_1 \|\widehat{\Sigma}' - \Sigma^*\|_{\max}$, where we use the fact that $\|AB\|_{\max} \leq \|A^\top\|_1 \|B\|_{\max}$ and $\Omega^*$ is symmetric. By Lemma E.5, we have $\|\widehat{\Sigma}' - \Sigma^*\|_{\max} = \mathcal{O}_\mathbb{P}(\sqrt{\log d/n})$. Under Assumption (**A5**) that $\|\Omega^*\|_1 = \mathcal{O}(1)$, we have $\|\Omega^* \widehat{\Sigma}' - I\|_{\max} = \mathcal{O}_\mathbb{P}(\sqrt{\log d/n})$. □

**Lemma F.2.** (Feasibility of $\Omega_{\mathcal{G}}^*$) Recall that $\widehat{\Sigma}_{\mathcal{G}}'$ is the diagonal block of $\widehat{\Sigma}'$ corresponding to $X_{\mathcal{G}_1}$ and $X_{\mathcal{G}_2}$, we have $\|\Omega_{\mathcal{G}}^* \widehat{\Sigma}_{\mathcal{G}}' - I\|_{\max} = \mathcal{O}_\mathbb{P}(\sqrt{\log d/n})$.

*Proof.* Since $\Omega_{\mathcal{G}}^* \widehat{\Sigma}_{\mathcal{G}}' - I = \Omega_{\mathcal{G}}^*(\widehat{\Sigma}_{\mathcal{G}}' - \Sigma_{\mathcal{G}}^*)$, we have $\|\Omega_{\mathcal{G}}^* \widehat{\Sigma}_{\mathcal{G}}' - I\|_{\max} \leq \|\Omega_{\mathcal{G}}^*\|_1 \|\widehat{\Sigma}_{\mathcal{G}}' - \Sigma_{\mathcal{G}}^*\|_{\max}$, where we use the fact that $\|AB\|_{\max} \leq \|A^\top\|_1 \|B\|_{\max}$ and $\Omega_{\mathcal{G}}^*$ is symmetric. By Lemma E.5, we have $\|\widehat{\Sigma}_{\mathcal{G}}' - \Sigma_{\mathcal{G}}^*\|_{\max} = \mathcal{O}_\mathbb{P}(\sqrt{\log d/n})$. For $\|\Omega_{\mathcal{G}}^*\|_1$, we have

$$\|\Omega_{\mathcal{G}}^*\|_1 = \|\Omega^* - \Theta^*\|_1 \leq \|\Omega^*\|_1 + \|\Theta^*\|_1 \leq \|\Omega^*\|_1 + \|\Theta^*\|_{1,1},$$

where we use the definition of $\Theta^*$, the triangle inequality and the fact that $\|A\|_1 \leq \|A\|_{1,1}$. Further under Assumption (**A5**) that $\|\Omega^*\|_1 = \mathcal{O}(1)$ and Assumption (**A3**) that $\|\Theta^*\|_{1,1} = \mathcal{O}(1)$, we have $\|\Omega_{\mathcal{G}}^*\|_1 = \mathcal{O}(1)$. Thus $\|\Omega_{\mathcal{G}}^* \widehat{\Sigma}'_{\mathcal{G}} - I\|_{\max} = \mathcal{O}_{\mathbb{P}}(\sqrt{\log d/n})$. $\square$

## F.1 Bias Correction Matrices

*Proof of Lemma D.1.* By Lemma F.1 and Lemma F.2, we can see that for $\lambda' = C'\sqrt{\log d/n}$, where $C'$ is a sufficiently large constant in (3.5) and (3.6), $\Omega^*$ and $\Omega_{\mathcal{G}}^*$ are feasible solutions to (3.5) and (3.6) with high probability. Since $M$ and $P$ minimizes the $\ell_\infty$ norm over the feasible solutions, we can easily get $\|P\|_\infty \leq \|\Omega_{\mathcal{G}}^*\|_1$ and $\|M\|_\infty \leq \|\Omega^*\|_1$. Thus we have $\|P\|_\infty = \mathcal{O}_{\mathbb{P}}(1)$ and $\|M\|_\infty = \mathcal{O}_{\mathbb{P}}(1)$. By triangle inequality, we have

$$\|M\widehat{\Sigma} - I\|_{\max} \leq \|M(\widehat{\Sigma} - \Sigma^*)\|_{\max} + \|M(\widehat{\Sigma}' - \Sigma^*)\|_{\max} + \|M\widehat{\Sigma}' - I\|_{\max}$$
$$\leq \|M\|_\infty \|\widehat{\Sigma} - \Sigma^*\|_{\max} + \|M\|_\infty \|\widehat{\Sigma}' - \Sigma^*\|_{\max} + \|M\widehat{\Sigma}' - I\|_{\max},$$

where in the last inequality we use $\|AB\|_{\max} \leq \|A\|_\infty \|B\|_{\max}$. Thus $\|M\widehat{\Sigma} - I\|_{\max} = \mathcal{O}_{\mathbb{P}}(\sqrt{\log d/n})$. Similar proofs can show that $\|P\widehat{\Sigma}_{\mathcal{G}} - I\|_{\max} = \mathcal{O}_{\mathbb{P}}(\sqrt{\log d/n})$. $\square$

## F.2 Bounding the Remainder Term

*Proof of Lemma D.2.* Recall from (3.4) that

$$\text{Remainder} = \underbrace{-M(\widehat{\Sigma} - \Sigma^*)\Theta^*(\widehat{\Sigma}_{\mathcal{G}} - \Sigma_{\mathcal{G}}^*)P^\top}_{I_1} + \underbrace{\widehat{\Theta} - \Theta^* - M\widehat{\Sigma}(\widehat{\Theta} - \Theta^*)\widehat{\Sigma}_{\mathcal{G}} P^\top}_{I_2}.$$

It can be decomposed into two terms. We bound each term separately. For $I_1$, we have

$$\|I_1\|_{\max} \leq \|M(\widehat{\Sigma} - \Sigma^*)\|_{\max} \|\Theta^*\|_{1,1} \|(\widehat{\Sigma}_{\mathcal{G}} - \Sigma_{\mathcal{G}}^*)P^\top\|_{\max}$$
$$\leq \|M\|_\infty \|\widehat{\Sigma} - \Sigma^*\|_{\max} \|\Theta^*\|_{1,1} \|\widehat{\Sigma}_{\mathcal{G}} - \Sigma_{\mathcal{G}}^*\|_{\max} \|P\|_\infty,$$

where in the first inequality we use $\|ABC\|_{\max} \leq \|A\|_{\max}\|B\|_{1,1}\|C\|_{\max}$ and in the second inequality we use $\|AB\|_{\max} \leq \|A\|_\infty \|B\|_{\max}$. Combining Assumption (**A3**), Lemma D.1 and Lemma E.5, we have $\|I_1\|_{\max} = \mathcal{O}_{\mathbb{P}}(\log d/n)$. For $I_2$, we have

$$I_2 = \widehat{\Theta} - \Theta^* - (M\widehat{\Sigma} - I + I)(\widehat{\Theta} - \Theta^*)(\widehat{\Sigma}_{\mathcal{G}} P^\top - I + I)$$
$$= -(M\widehat{\Sigma} - I)(\widehat{\Theta} - \Theta^*)(\widehat{\Sigma}_{\mathcal{G}} P^\top - I) - (M\widehat{\Sigma} - I)(\widehat{\Theta} - \Theta^*) - (\widehat{\Theta} - \Theta^*)(\widehat{\Sigma}_{\mathcal{G}} P^\top - I).$$

Using the fact that $\|ABC\|_{\max} \leq \|A\|_{\max}\|B\|_{1,1}\|C\|_{\max}$, $\|AB\|_{\max} \leq \|A\|_{\max}\|B\|_{1,1}$ and $\|AB\|_{\max} \leq \|A\|_{1,1}\|B\|_{\max}$, we have

$$\|I_2\|_{\max} \leq \|M\widehat{\Sigma} - I\|_{\max} \|\widehat{\Theta} - \Theta^*\|_{1,1} \|\widehat{\Sigma}_{\mathcal{G}} P^\top - I\|_{\max} + \|M\widehat{\Sigma} - I\|_{\max} \|\widehat{\Theta} - \Theta^*\|_{1,1}$$
$$+ \|\widehat{\Theta} - \Theta^*\|_{1,1} \|\widehat{\Sigma}_{\mathcal{G}} P^\top - I\|_{\max}.$$

By Theorem 4.1 and Lemma D.1, we have $\|I_2\|_{\max} = \mathcal{O}_{\mathbb{P}}(s^* \log d/n)$. Combining with the fact that $\|I_1\|_{\max} = \mathcal{O}_{\mathbb{P}}(\log d/n)$, we have $\|\text{Remainder}\|_{\max} = \mathcal{O}_{\mathbb{P}}(s^* \log d/n)$. $\square$

## F.3 Variance Lower Bound

*Proof of Lemma D.3.* Recall from (4.1), we have

$$\xi_{jk}^2 = \underbrace{(M_{j*}\Sigma^* M_{j*}^\top)[P_{k*}(I + \Sigma_{\mathcal{G}}^*\Theta^*)\Sigma_{\mathcal{G}}^* P_{k*}^\top]}_{I_1} + \underbrace{(M_{j*}\Sigma_{\mathcal{G}}^* P_{k*}^\top)^2}_{I_2} - \underbrace{(M_{j*}\Sigma^* P_{k*}^\top)^2}_{I_3}$$
$$- \underbrace{[M_{j*}(I - \Sigma^*\Theta^*)\Sigma_{\mathcal{G}_2}^*(I - \Theta^*\Sigma^*)M_{j*}^\top](P_{k*}\Sigma_{\mathcal{G}}^* P_{k*}^\top)}_{I_4}. \tag{F.1}$$

It is equivalent to showing that with high probability $\xi_{jk}^2$ is lower bounded by a constant. To this end, we treat the four terms in (F.1) separately. For $I_1$, we aim to show that $I_1 \geq (1 - \lambda')^4/(\sigma_{jj}^* \sigma_{kk}^*)$, where $\lambda'$ is same constant as in (3.5) and (3.6). For $I_2$, we use the lower bound $I_2 \geq 0$. For the remaining two $I_3$ and $I_4$, we aim to show that they are $o_{\mathbb{P}}(1)$.

We first show that $I_1$ is lower bounded by a constant. Due to the constraint in (3.5), we have $|1 - M_{j*}\widehat{\Sigma}' e_j| \leq \lambda'$, where $e_j$ is the $j$-th natural basis in $\mathbb{R}^d$. By the proof of Lemma D.1, we also have $|1 - M_{j*}\Sigma^* e_j| = \mathcal{O}_{\mathbb{P}}(\lambda')$. Consider the following convex optimization problem

$$\min_{v \in \mathbb{R}^d} \quad v^\top \Sigma^* v$$
$$\text{subject to} \quad 1 - v^\top \Sigma^* e_j \leq \lambda'. \tag{F.2}$$

We can see that with high probability $v = M_{j*}^\top$ is a feasible solution to (F.2). To lower bound $M_{j*}\Sigma^* M_{j*}^\top$, we consider the dual problem of (F.2), which is given by

$$\max_{c \in \mathbb{R}} \quad c(1 - \lambda') - \frac{c^2}{4}\sigma_{jj}^*$$
$$\text{subject to} \quad c \geq 0. \tag{F.3}$$

The optimal value of (F.3) is $(1 - \lambda')^2/\sigma_{jj}^*$. Thus by weak duality, we have for any feasible $v$ of (F.2), $v^\top \Sigma^* v \geq (1 - \lambda')^2/\sigma_{jj}^*$. Since $v = M_{j*}^\top$ is a feasible solution, we have $M_{j*}\Sigma^* M_{j*}^\top \geq (1 - \lambda')^2/\sigma_{jj}^*$. Similarly we have $P_{k*}(I + \Sigma_{\mathcal{G}}^*\Theta^*)\Sigma_{\mathcal{G}}^* P_{k*}^\top \geq (1 - \lambda')^2/\sigma_{kk}^*$.

For $I_3$, we have $(M_{j*}\Sigma^* P_{k*}^\top)^2 = [(M_{j*}\Sigma^* - e_j^\top)P_{k*}^\top]^2$, where we use the fact that $e_j^\top P_{k*}^\top = 0$. This is the case since $P$ has a block structure and $j, k$ are off diagonal indexes. Further by Cauchy Schwarz inequality, we have $I_3 \leq \|M_{j*}\Sigma^* - e_j^\top\|_\infty^2 \|P_{k*}\|_1^2 = \mathcal{O}_{\mathbb{P}}((\lambda')^2)$, where we use the fact that $\|P_{k*}\|_1 \leq \|P\|_\infty = \mathcal{O}_{\mathbb{P}}(1)$ and $\|M_{j*}\Sigma^* - e_j^\top\|_\infty \leq \|M\Sigma^* - I\|_{\max} = \mathcal{O}_{\mathbb{P}}(\lambda')$.

Finally for $I_4$, we have $|P_{k*}\Sigma_{\mathcal{G}}^* P_{k*}^\top| \leq \|P_{k*}\|_1^2 \|\Sigma_{\mathcal{G}}^*\|_{\max} = \mathcal{O}_{\mathbb{P}}(1)$ due to Assumption (**A2**) that $\|\Sigma_{\mathcal{G}}^*\| = \mathcal{O}(1)$ and the fact that $\|P_{k*}\|_1 = \mathcal{O}_{\mathbb{P}}(1)$. For the other term in $I_4$, we have

$$M_{j*}(I - \Sigma^*\Theta^*)\Sigma_{\mathcal{G}_2}^*(I - \Theta^*\Sigma^*)M_{j*}^\top = M_{j*}\Sigma^* I_2(I - \Theta^*\Sigma^*)M_{j*}^\top$$
$$= (M_{j*} - \Omega_{j*}^*)\Sigma^* I_2(I - \Theta^*\Sigma^*)M_{j*}^\top. \tag{F.4}$$

where in the first equality $I_2 = \text{diag}(0, I_{d_2})$ and in the second equality we use the fact that $\Omega_{j*}^*\Sigma^* I_2 = e_j^\top I_2 = 0$. This is true because $j \in [d_1]$. By Cauchy Schwarz inequality, we have

$$|(M_{j*} - \Omega_{j*}^*)\Sigma^* I_2(I - \Theta^*\Sigma^*)M_{j*}^\top| \leq \|(M_{j*} - \Omega_{j*}^*)\Sigma^* I_2(I - \Theta^*\Sigma^*)\|_\infty \|M_{j*}^\top\|_1. \tag{F.5}$$

We already know that $\|M_{j*}^\top\|_1 = \mathcal{O}_\mathbb{P}(1)$. To tackle the other term in (F.5), we use the fact that $\|AB\|_{\max} \leq \|A\|_{\max}\|B\|_1$, then we get

$$\|(M_{j*} - \Omega_{j*}^*)\Sigma^* I_2(I - \Theta^*\Sigma^*)\|_\infty \leq \|M\Sigma^* - I\|_{\max}\|I_2(I - \Theta^*\Sigma^*)\|_1$$
$$\leq \|M\Sigma^* - I\|_{\max}\|I - \Theta^*\Sigma^*\|_1, \tag{F.6}$$

where in the first inequality we use the fact that $\|(M_{j*} - \Omega_{j*}^*)\Sigma^*\|_{\max} \leq \|M\Sigma^* - I\|_{\max}$. By the proof of Lemma D.1, we have $\|M\Sigma^* - I\|_{\max} = \mathcal{O}_\mathbb{P}(\lambda')$. And from (E.7), we have $\|I - \Theta^*\Sigma^*\|_1 = \mathcal{O}(1)$. Combining (F.4), (F.5) and (F.6), we have $|M_{j*}(I - \Sigma^*\Theta^*)\Sigma_{\mathcal{G}_2}^*(I - \Theta^*\Sigma^*)M_{j*}^\top| = \mathcal{O}_\mathbb{P}(\lambda')$. Further because $|P_{k*}\Sigma_\mathcal{G}^* P_{k*}^\top| = \mathcal{O}_\mathbb{P}(1)$, we have $I_4 = \mathcal{O}_\mathbb{P}(\lambda')$.

Combining all, we have with high probability

$$\xi_{jk}^2 \geq \frac{(1-\lambda')^4}{\sigma_{jj}^*\sigma_{kk}^*} - (\lambda')^2 - \lambda' \geq \frac{(1-\lambda')^4}{2\sigma_{jj}^*\sigma_{kk}^*},$$

where we use the fact that $\lambda' \asymp \sqrt{\log d/n}$ and $\log d/n = o(1)$. In all, we show that with high probability, $\xi_{jk}^2$ is lower bounded by a constant, thus $1/\xi_{jk} = \mathcal{O}_\mathbb{P}(1)$. □

## F.4 Tail Bound

*Proof of Lemma D.5.* Define four random variables $Y_1 = -u^\top X, Y_2 = X^\top(I + \Theta^*\Sigma_\mathcal{G}^*)v, Y_3 = u^\top(I - \Sigma^*\Theta^*)X_{\mathcal{G}_2}$ and $Y_4 = X_{\mathcal{G}_2}^\top v$. Hence $Z = Y_1Y_2 - \mathbb{E}[Y_1Y_2] + Y_3Y_4 - \mathbb{E}[Y_3Y_4]$. From the definitions, we have that $Y_1 \sim N(0, u^\top\Sigma^*u), Y_2 \sim N(0, v^\top\Sigma_\mathcal{G}^*(I + \Theta^*\Sigma_\mathcal{G}^*)v), Y_3 \sim N(0, u^\top(I - \Sigma^*\Theta^*)\Sigma_2^*(I - \Theta^*\Sigma^*)u)$ and $Y_4 \sim N(0, v^\top\Sigma_2^*v)$. It is easy to show that they all have finite variance. Thus for $i \in [4]$, $\|Y_i\|_{\psi_2} = \mathcal{O}(1)$. By Lemma E.3, we have $\|Y_1Y_2\|_{\psi_1} = \mathcal{O}(1)$ and $\|Y_3Y_4\|_{\psi_1} = \mathcal{O}(1)$. Thus $\|Z\|_{\psi_1} = \mathcal{O}(1)$ by triangle inequality and the centering property of the sub-exponential norm. □